\documentclass[11pt]{article}

\newif\ifArxivVersion
\ArxivVersiontrue % arXiv: use the saved bibliography in bib.bbl.
\usepackage[margin=0.9in]{geometry}
\usepackage[T1]{fontenc}
\usepackage[utf8]{inputenc}
\usepackage{tgheros}

\usepackage{mathtools}
\mathtoolsset{showonlyrefs}
\usepackage{amssymb}
\usepackage{booktabs}
\usepackage{enumitem}
\usepackage{microtype}
\usepackage{bm}
\usepackage[numbers,square,sort&compress]{natbib}
\usepackage[dvipsnames]{xcolor}
\usepackage{tabularx}
\usepackage{graphicx}
\usepackage{caption}
\usepackage{subcaption}
\usepackage{float}
\usepackage{placeins}
\usepackage{pdflscape}

\usepackage{multirow}

\usepackage{ragged2e}

\graphicspath{{Plots/}}

\floatstyle{plaintop}
\newfloat{algorithm}{tbp}{loa}
\floatname{algorithm}{ALGORITHM}

\usepackage[
  colorlinks = true,
  linkcolor = blue,
  citecolor = blue,
  urlcolor = blue,
  hypertexnames = false
]{hyperref}
\usepackage{doi}

\usepackage{orcidlink}

\renewcommand{\sectionautorefname}{Section}
\renewcommand{\subsectionautorefname}{Section}
\renewcommand{\subsubsectionautorefname}{Section}
\providecommand{\algorithmautorefname}{Algorithm}

\newcommand{\mainsecref}[1]{\hyperref[#1]{Section~\ref*{#1}}}
\newcommand{\mainfigref}[1]{\hyperref[#1]{Figure~\ref*{#1}}}
\newcommand{\maintabref}[1]{\hyperref[#1]{Table~\ref*{#1}}}
\newcommand{\mainalgref}[1]{\hyperref[#1]{Algorithm~\ref*{#1}}}

\newcommand{\appref}[1]{\hyperref[#1]{Appendix~\ref*{#1}}}
\newcommand{\apptabref}[1]{\hyperref[#1]{Appendix Table~\ref*{#1}}}
\newcommand{\appfigref}[1]{\hyperref[#1]{Appendix Figure~\ref*{#1}}}
\newcommand{\appalgref}[1]{\hyperref[#1]{Algorithm~\ref*{#1}}}

\newcommand{\suppsecref}[1]{\hyperref[#1]{Supplementary Section~\ref*{#1}}}
\newcommand{\supptabref}[1]{\hyperref[#1]{Supplementary Table~\ref*{#1}}}
\newcommand{\suppfigref}[1]{\hyperref[#1]{Supplementary Figure~\ref*{#1}}}
\newcommand{\suppalgref}[1]{\hyperref[#1]{Supplementary Algorithm~\ref*{#1}}}
\newcommand{\suppeqref}[1]{\hyperref[#1]{Supplementary Equation~\ref*{#1}}}

\newcommand{\E}{\mathbb{E}}
\newcommand{\Var}{\operatorname{Var}}

\newcommand{\Normal}{\mathcal{N}}
\newcommand{\NB}{\mathrm{NB}}

\newcommand{\Nzero}{\mathbb{N}_{0}}
\newcommand{\R}{\mathbb{R}}
\newcommand{\obs}{\mathcal{O}}
\newcommand{\mis}{\mathcal{M}}
\newcommand{\obsval}{\mathrm{obs}}
\newcommand{\pre}{\mathrm{pre}}
\newcommand{\post}{\mathrm{post}}
\newcommand{\comp}{\mathrm{comp}}
\newcommand{\misval}{\mathrm{mis}}
\newcommand{\asgn}{\mathrm{asgn}}
\newcommand{\refarm}{\mathrm{ref}}
\newcommand{\aug}{\mathrm{aug}}

\newcommand{\ManuscriptTitle}{Reference-based sensitivity analysis for repeated count outcomes using negative binomial margins and a Gaussian copula}

\newcommand{\AuthorBlock}{
Divan A. Burger\,\orcidlink{0000-0001-8096-6371}$^{1,2,*}$,
Emmanuel Lesaffre\,\orcidlink{0000-0002-3747-6905}$^{3,4}$,
Reynaldo Martina\,\orcidlink{0000-0002-1925-820X}$^{5}$
}

\newcommand{\AffiliationBlock}{
\begin{tabularx}{\textwidth}{@{}r@{\hspace{0.4em}}X@{}}
$^{1}$ & Cytel Inc., Waltham, MA, USA\\
$^{2}$ & Department of Mathematical Statistics and Actuarial Science, University of the Free State, Bloemfontein, South Africa\\
$^{3}$ & L-BioStat, KU Leuven, Leuven, Belgium\\
$^{4}$ & Department of Statistics and Actuarial Science, Stellenbosch University, Stellenbosch, South Africa\\
$^{5}$ & Department of Population Health Sciences, Faculty of Veterinary Medicine, Utrecht University, Utrecht, The Netherlands
\end{tabularx}
}

\newcommand{\CorrespondingAuthor}{
\begin{tabularx}{\textwidth}{@{}r@{\hspace{0.27em}}X@{}}
$^{*}$ & Corresponding author: Divan A. Burger, Cytel Inc., 1050 Winter Street, Waltham, MA 02451, USA. Email: \href{mailto:divanaburger@gmail.com}{\texttt{divanaburger@gmail.com}}
\end{tabularx}
}

\newcommand{\MainAbstractText}{
Reference-based multiple imputation is used in longitudinal clinical trials to assess sensitivity to assumptions about outcomes unobserved after intercurrent events. Most existing methods target continuous outcomes and use multivariate normal working models. Scheduled count outcomes require a model that preserves integer support, skewness, overdispersion, longitudinal dependence, and an exposure-based rate interpretation. We propose visit-specific negative binomial (NB) margins linked by a Gaussian copula. Covariate-adjusted log rate models define the margins, while the copula captures longitudinal dependence. For missing outcomes in the active arm after a prespecified event, termed the trigger, assigned arm continuation retains the fitted active arm marginal mean, jump to reference uses the corresponding reference arm marginal mean, and the intermediate rule interpolates between these means on the log rate scale. Combined with the copula, the resulting NB margins determine an imputation distribution conditional on the retained history. We account for discreteness through randomized probability integral transform augmentation and fit the model with a custom Metropolis-within-Gibbs sampler. In targeted simulations, the model recovered the generating marginal and dependence parameters with little bias. In simulations with incomplete data, treatment effect estimates were closest to the corresponding complete data estimates when the imputation rule matched the mechanism governing outcomes after the trigger. We illustrate the method using repeated incontinence episode counts from a published trial in overactive bladder. Estimated rate ratios comparing active treatment with placebo remained below 1 under all reference-based assumptions, with modest attenuation toward the null and the greatest separation between rules at the final visit.
}

\newcommand{\KeywordBlock}{
\begin{tabularx}{\textwidth}{@{}l@{\hspace{0.27em}}X@{}}
\textbf{Keywords:} & Reference-based multiple imputation; Missing data; Count outcomes; Negative binomial model; Gaussian copula
\end{tabularx}
}

\newcommand{\MainTitlePageContent}{
\noindent\textbf{Abstract}

\vspace{0.25cm}

\noindent\ignorespaces\MainAbstractText

\vspace{0.4cm}

\KeywordBlock
}

\newcommand{\SupplementTitle}{
Supplementary material for:\\[0.35em]
\ManuscriptTitle
}

\newcommand{\IfNotEmpty}[2]{
\if\relax\detokenize{#1}\relax
\else
#2
\fi
}

\newcommand{\MakeManuscriptTitlePage}[2]{
\begin{titlepage}
\thispagestyle{empty}

{\Large\bfseries #1\par}

\vspace{0.7cm}

\rule{\textwidth}{0.4pt}

\vspace{0.7cm}

{\large \AuthorBlock\par}

\vspace{0.55cm}

\begin{minipage}{\textwidth}
{\small \AffiliationBlock\par}
\end{minipage}

\vspace{0.45cm}

\begin{minipage}{\textwidth}
{\normalsize \CorrespondingAuthor\par}
\end{minipage}

\vspace{0.7cm}

\IfNotEmpty{#2}{
\begin{minipage}{\textwidth}
#2
\end{minipage}
}

\end{titlepage}

\newpage
}

\newcommand{\StartSupplement}{
\clearpage

\MakeManuscriptTitlePage{\SupplementTitle}{}

\setcounter{section}{0}
\setcounter{subsection}{0}
\setcounter{subsubsection}{0}
\setcounter{figure}{0}
\setcounter{table}{0}
\setcounter{algorithm}{0}
\setcounter{equation}{0}

\renewcommand{\thesection}{S\arabic{section}}
\renewcommand{\thesubsection}{S\arabic{section}.\arabic{subsection}}
\renewcommand{\thesubsubsection}{S\arabic{section}.\arabic{subsection}.\arabic{subsubsection}}
\renewcommand{\thefigure}{S\arabic{figure}}
\renewcommand{\thetable}{S\arabic{table}}
\renewcommand{\thealgorithm}{S\arabic{algorithm}}
\renewcommand{\theequation}{S\arabic{equation}}

\renewcommand{\theHsection}{supp.\arabic{section}}
\renewcommand{\theHsubsection}{supp.\arabic{section}.\arabic{subsection}}
\renewcommand{\theHsubsubsection}{supp.\arabic{section}.\arabic{subsection}.\arabic{subsubsection}}
\renewcommand{\theHfigure}{supp.\arabic{figure}}
\renewcommand{\theHtable}{supp.\arabic{table}}
\providecommand{\theHalgorithm}{\thealgorithm}
\renewcommand{\theHalgorithm}{supp.\arabic{algorithm}}
\renewcommand{\theHequation}{supp.\arabic{equation}}

\renewcommand{\sectionautorefname}{Supplementary Section}
\renewcommand{\subsectionautorefname}{Supplementary Section}
\renewcommand{\subsubsectionautorefname}{Supplementary Section}
\renewcommand{\figureautorefname}{Supplementary Figure}
\renewcommand{\tableautorefname}{Supplementary Table}
\renewcommand{\equationautorefname}{Supplementary Equation}
\renewcommand{\algorithmautorefname}{Supplementary Algorithm}
}

\begin{document}

\MakeManuscriptTitlePage{\ManuscriptTitle}{\MainTitlePageContent}

\section{Introduction}

In longitudinal clinical trials, reference-based multiple imputation (RBMI) is commonly used to assess how conclusions change under alternative assumptions about outcomes that are unobserved after treatment discontinuation, rescue therapy, or other intercurrent events \citep{carpenter2013analysis,cro2020sensitivity}. Most practical implementations have been developed for continuous endpoints. For such endpoints, a multivariate normal working model for the scheduled outcome vector defines both the joint distribution of the repeated measurements and the conditional distributions used to impute missing outcomes given the retained history \citep{rubin1987,carpenter2013analysis}. Assigned arm continuation imputes missing outcomes from the distribution for the randomized arm. Jump to reference (J2R) instead uses the corresponding reference arm marginal mean trajectory for missing outcomes in the active arm after the intercurrent event. In both cases, the fitted longitudinal model conditions the imputations on the retained history. Intermediate controlled rules retain a prespecified proportion of the treatment effect; this proportion can be varied as a sensitivity parameter or in a tipping point analysis \citep{cro2020sensitivity,white2020causal}.

Recent work has extended RBMI to longitudinal binary outcomes through a latent normal model \citep{cro2025reference}. Scheduled counts pose a different problem. Their imputation model should preserve integer support, skewness, overdispersion, and, where relevant, a rate interpretation based on exposure offsets. A multivariate normal model applied directly to counts does not preserve these features. Negative binomial (NB) regression gives a natural marginal model for overdispersed counts, but separate models at each visit do not by themselves define a longitudinal joint distribution or the conditional distributions needed to impute missing counts from retained observations. A suitable method must therefore preserve NB margins at the scheduled visits while also supporting conditional imputation under reference-based assumptions.

Several methods address related count settings. Controlled multiple imputation methods, including reference-based approaches, have been developed for recurrent event endpoints under a range of count and event process models \citep{keene2014missing,akacha2016sensitivity,gao2017control,diao2022efficient}; the approach of Keene et al.~\citep{keene2014missing} is implemented in the R package \texttt{dejaVu} \citep{burkoff2024dejavu}. These approaches concern event counts accumulated over follow-up or recurrent event processes, rather than a joint vector of counts recorded at scheduled visits. For multivariate nonnormal longitudinal outcomes, Tang \citep{tang2019monotone} proposed sequential conditional regression models, including Poisson and NB models, that support controlled imputation under nonignorable dropout. Although developed mainly for the design and simulation of trials with NB outcomes, version 0.3.2 of the R package \texttt{gsDesignNB} includes a multiple imputation module for longitudinal NB counts. Its copy-reference option combines fixed-effect predictions from the reference arm with a multiplier based on subject-specific random effects \citep{anderson2026gsdesignnb}. These methods address related count imputation problems but do not provide the marginal copula construction developed here for repeated counts at scheduled visits.

We link visit-specific NB margins with a Gaussian copula. This allows the assumptions about marginal outcomes after the trigger to be specified separately from the dependence model that conditions imputations on the retained history. Gaussian copulas have previously been used to impute mixed or nonnormal data \citep{hollenbach2021multiple,zhao2020missing,zhao2024gcimpute}, and marginal regression with Gaussian copulas has been developed for time series, longitudinal, and spatial data, including longitudinal NB counts \citep{masarotto2017gaussian}. We build on this work to obtain reference-based conditional imputation distributions for counts recorded at scheduled visits.

The International Council for Harmonisation (ICH) E9(R1) addendum distinguishes intercurrent events from missing data and treats the strategy for handling intercurrent events as part of the estimand \citep{ich2019e9r1}. In our approach, the trigger may be a prespecified intercurrent event or an operational trigger defined for the analysis. Recorded counts before the trigger form the retained conditioning set, and scheduled outcomes from the trigger onward are imputed under a prespecified sensitivity rule. For participants in the active arm, each rule specifies the marginal log mean after the trigger, while the fitted copula conditions their imputations on recorded counts before the trigger. All rules are identical for participants in the reference arm.

The discreteness of the margins requires special care. An observed count identifies an interval, rather than a single value, on the latent Gaussian scale. We handle this using randomized probability integral transform (PIT) augmentation, following data augmentation methods for copula models with discrete margins \citep{smith2012estimation}. Auxiliary variables locate the observed counts within their probability intervals and are sampled jointly with the model parameters. Posterior inference therefore averages over the uncertainty in these locations instead of fixing arbitrary transformed values.

Our contribution has three parts. First, we express reference-based assumptions after the trigger through covariate-adjusted NB margins at each visit. The marginal models include exposure offsets and allow a continuously varying proportion of the treatment effect to be retained on the log rate scale. Second, a Gaussian copula links these margins, giving a longitudinal joint distribution and conditional imputation distributions for any pattern of retained visits. Third, we provide a Bayesian implementation using randomized PIT augmentation and a Metropolis-within-Gibbs sampler. This implementation carries uncertainty in the marginal parameters, dependence structure, and latent locations of the retained counts into the imputations.

\autoref{sec:motivating} describes the motivating trial, and \autoref{sec:methods} presents the marginal model, copula, reference-based rules, imputation algorithm, and estimation method. \autoref{sec:simulation} reports one study of model recovery and a second comparing treatment effect estimates after imputation with estimates from the corresponding complete data before missingness was imposed. We apply the method in \autoref{sec:application} and discuss its interpretation and possible extensions in \autoref{sec:discussion}. Supporting derivations and implementation details are provided in the appendices, with additional simulation and application results in the supplementary material.

\section{Motivating clinical trial}
\label{sec:motivating}

The motivating data come from a 12-week, multicenter, double-blind Phase~3 trial with placebo and active controls, conducted in Europe and Australia (ClinicalTrials.gov identifier NCT00689104). Patients were randomized to once-daily placebo, mirabegron 50~mg, mirabegron 100~mg, or tolterodine extended-release 4~mg \citep{khullar2013phase3}. The co-primary efficacy endpoints were the changes from baseline to the final visit in the mean numbers of incontinence episodes and micturitions per 24 hours. Our application uses the placebo and mirabegron 50~mg groups and focuses on repeated counts of incontinence episodes.

For this methodological illustration, the available application dataset comprised participants from the trial's full analysis set for incontinence (FAS-I) in the placebo and mirabegron 50~mg groups. In the original trial, FAS-I consisted of patients in the full analysis set who had at least one incontinence episode at baseline \citep{khullar2013phase3}. We required a recorded baseline count, positive baseline diary exposure, and complete data for the baseline covariates used in the imputation and analysis models. Patients were not excluded because a later post-baseline count was missing. The resulting sample contained 584 patients, of whom 291 received placebo and 293 received mirabegron 50~mg. Counts of incontinence episodes were recorded at baseline and at Weeks~4, 8, and 12 over diary windows. For each count, exposure was defined as the number of valid diary days and varied for some patients and visits.

\begin{figure}[!htbp]
\centering
\includegraphics[width=0.92\textwidth]{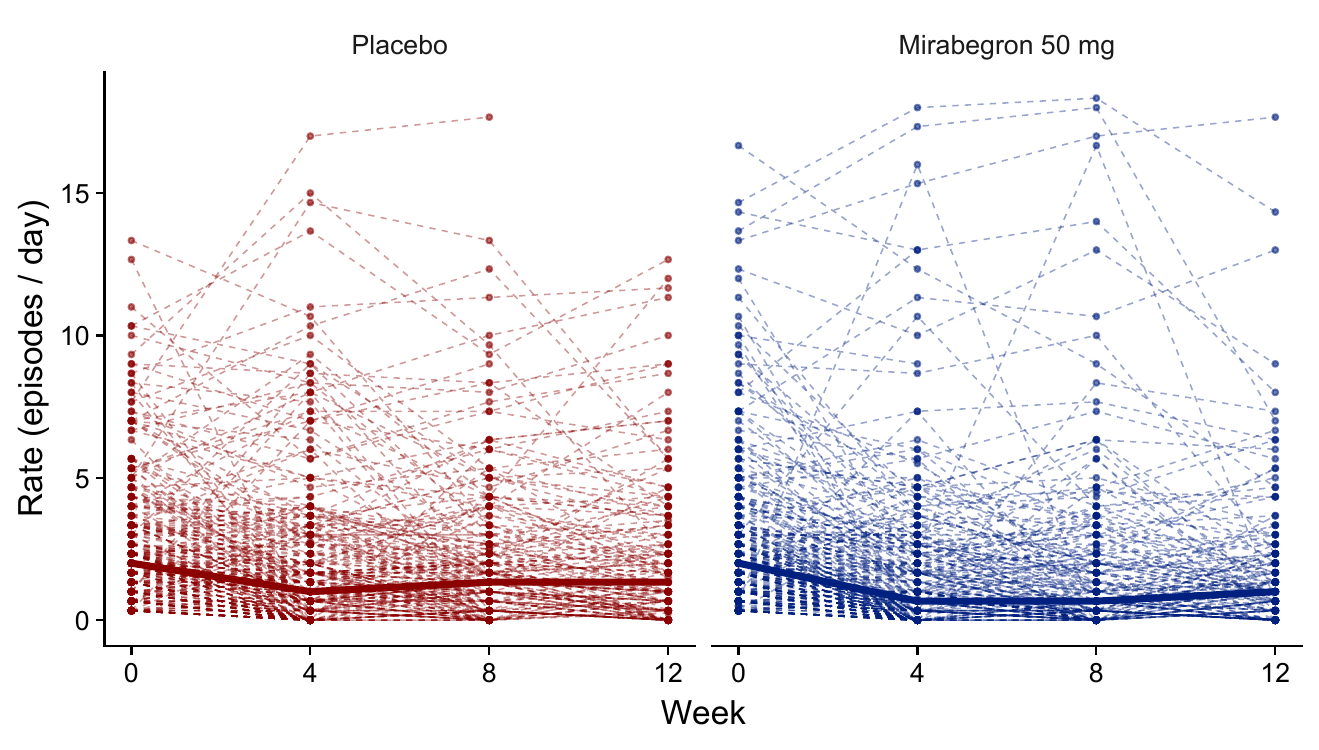}
\caption{
Observed daily rates of incontinence episodes by treatment group.
Thin dashed lines represent individual patients, and the thicker solid line gives the median for each treatment group at each visit. Daily rates were calculated by dividing the number of incontinence episodes by the number of valid diary days.
}
\label{fig:motivating-rate-profiles}
\end{figure}

\begin{figure}[!htbp]
\centering
\begin{subfigure}[t]{0.49\textwidth}
\centering
\includegraphics[width=\textwidth]{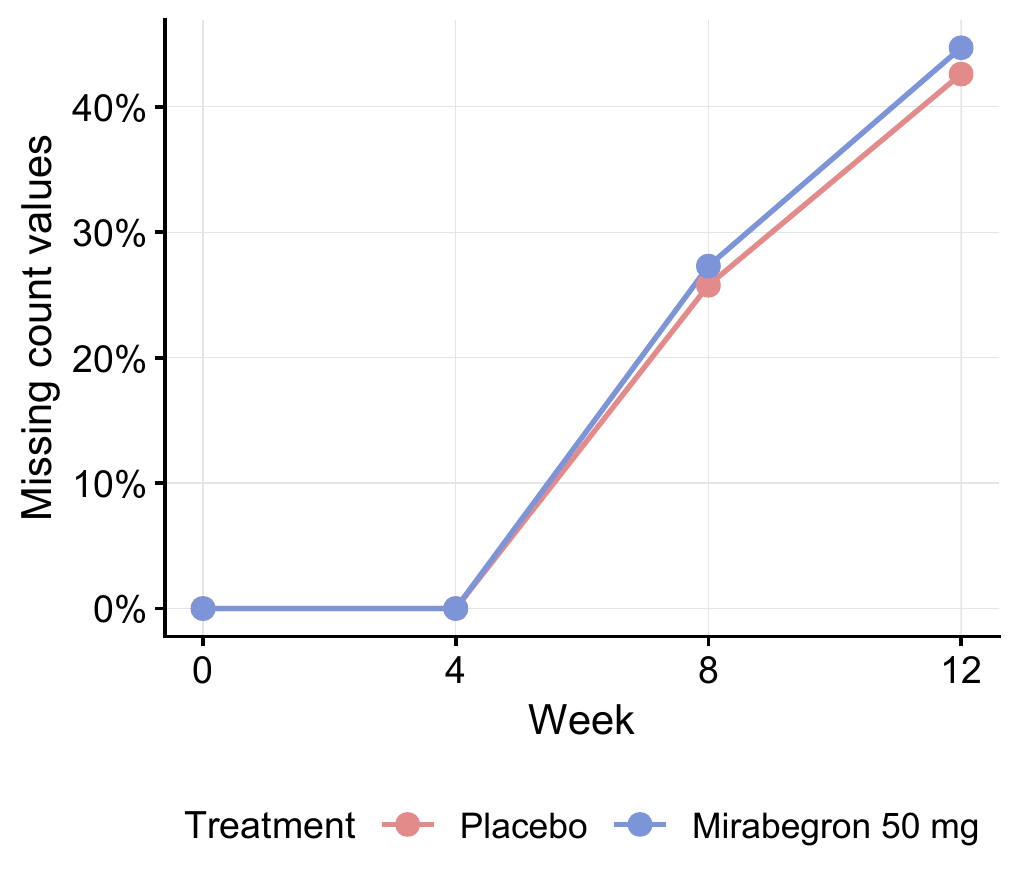}
\caption{Missingness by visit.}
\label{fig:motivating-missingness-panel}
\end{subfigure}
\hfill
\begin{subfigure}[t]{0.49\textwidth}
\centering
\includegraphics[width=\textwidth]{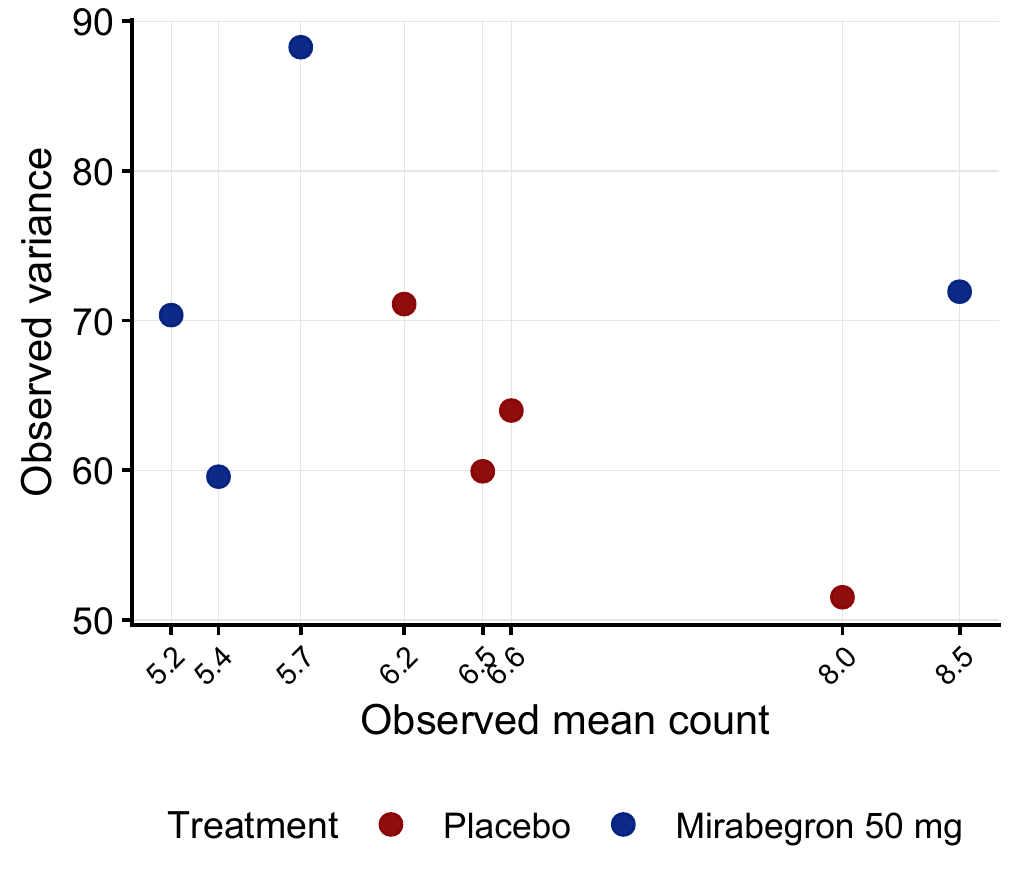}
\caption{Mean-variance relationship.}
\label{fig:motivating-meanvariance-panel}
\end{subfigure}
\caption{
Missingness and overdispersion in the motivating data.
Panel~(a) shows the increase in missingness over follow-up. Panel~(b) compares the observed mean and variance of the incontinence episode counts by visit and treatment group; the variances are substantially larger than the corresponding means.
}
\label{fig:motivating-missingness-overdispersion}
\end{figure}

Several features of the observed data motivate the method. The endpoint is a nonnegative integer count accumulated over a known observation period, so it has a natural rate interpretation. Individual daily rates varied considerably over follow-up (\autoref{fig:motivating-rate-profiles}). The counts were also overdispersed. At baseline, the mean counts over the diary window were 8.0 with placebo and 8.5 with mirabegron 50~mg, whereas the corresponding variances were 51.5 and 71.9. Overdispersion remained evident after baseline in the mean-variance panel of \autoref{fig:motivating-missingness-overdispersion}. Missingness increased monotonically over follow-up, with all baseline and Week~4 counts recorded. Patients either had counts through Week~12, were missing only at Week~12, or were missing from Week~8 onward. These patterns occurred in 167 (57.4\%), 49 (16.8\%), and 75 (25.8\%) placebo patients, respectively, and in 162 (55.3\%), 51 (17.4\%), and 80 (27.3\%) patients receiving mirabegron 50~mg.

Across treatment groups and post-baseline visits, 17.1\% to 32.1\% of the recorded counts were zero. A high proportion of zeros does not by itself imply zero inflation \citep{warton2005many}, so we assessed excess zeros with a separate working NB mixed model fitted to the observed post-baseline counts. This model included fixed effects for treatment by visit, an offset for log exposure, visit-specific overdispersion parameters, and a subject-specific random intercept \citep{brooks2017glmmtmb}. A diagnostic based on 2,000 simulations conditional on the fitted random effects gave an observed-to-expected zero ratio of 1.089 and a $p$-value of 0.147. This provided no evidence of more zeros than the NB model could accommodate \citep{HARTIG2021A}.

Reliable indicators of treatment discontinuation and rescue medication use were unavailable in the application dataset. We therefore used the first missing scheduled post-baseline count as the operational trigger for the reference-based rule. The combination of overdispersion, varying diary observation periods, and monotone missingness makes these data well suited for illustrating the proposed method.

\section{Methods}
\label{sec:methods}

\subsection{Data structure and conditioning sets defined by the trigger}

Consider a randomized trial with participants $i = 1, \ldots, n$ and scheduled post-baseline visits $j = 1, \ldots, J$. Let $A_i \in \left\{0,1\right\}$ denote the randomized treatment assignment for participant $i$, with $A_i = 0$ for the reference arm and $A_i = 1$ for the active arm. Let $X_i \in \R^p$ denote the column vector of baseline covariates for participant $i$.

Let
\begin{equation}
\Nzero
=
\left\{0,1,2,\ldots\right\}
\end{equation}
denote the set of nonnegative integers. For participant $i$, let
\begin{equation}
Y_i
=
\left(
Y_{i1},\ldots,Y_{iJ}
\right)^\top
\end{equation}
denote the full longitudinal count vector, where $Y_{ij} \in \Nzero$ is the scheduled count at visit $j$. Zero counts are therefore allowed. We use lowercase $y$ for realized values of the count variables.

Let $t_{ij}^{Y}$ be the prespecified analysis time for the scheduled assessment corresponding to $Y_{ij}$, ordered such that
\begin{equation}
t_{i1}^{Y}
<
\cdots
<
t_{iJ}^{Y}.
\end{equation}
Let $e_{ij} > 0$ denote the exposure used in the marginal model for $Y_{ij}$. For a recorded count, this is the observed duration over which the count accumulated. For a count that must be imputed, $e_{ij}$ is fixed according to a prespecified analysis convention for the intended observation window. Let
\begin{equation}
e_i
=
\left(
e_{i1},\ldots,e_{iJ}
\right)^\top
\end{equation}
denote the exposure vector. The imputation model conditions on $e_i$, with $\log\left(e_{ij}\right)$ included as an offset in the marginal model.

Let $O_{ij}$ denote the recording indicator before application of the trigger rule:
\begin{equation}
O_{ij}
=
\begin{cases}
1, & \text{if $Y_{ij}$ is recorded}, \\
0, & \text{otherwise}.
\end{cases}
\end{equation}
Let $T_i$ denote the time of the prespecified trigger, if one occurs. The trigger may be an intercurrent event or an operational trigger based on missing data. The first affected scheduled visit is
\begin{equation}
D_i
=
\begin{cases}
\min\left\{j:t_{ij}^{Y}\geq T_i\right\},
&
\text{if $T_i$ is defined and $T_i\leq t_{iJ}^{Y}$},
\\
J+1,
&
\text{otherwise}.
\end{cases}
\end{equation}
The value $D_i=J+1$ indicates that the trigger affects no scheduled post-baseline visit. Because the definition uses $\geq$, a scheduled assessment assigned the same analysis time as the trigger is considered affected.

For the reference-based sensitivity analysis, recorded values before $D_i$ are retained for conditioning. Any recorded values on or after $D_i$ are set aside and treated as requiring imputation. Define the set of conditioning visits as
\begin{equation}
\obs_i
=
\left\{
j \in \left\{1,\ldots,J\right\}:
O_{ij}=1 \text{ and } j<D_i
\right\}
\end{equation}
and the set of visits requiring imputation as
\begin{equation}
\mis_i
=
\left\{1,\ldots,J\right\}
\setminus
\obs_i.
\end{equation}
The set $\mis_i$ therefore contains both unrecorded measurements and recorded measurements set aside because they occur from the trigger onward.

Partition the visits requiring imputation as
\begin{equation}
\mis_i^{\pre}
=
\left\{
j \in \mis_i:j<D_i
\right\}
\end{equation}
and
\begin{equation}
\mis_i^{\post}
=
\left\{
j \in \mis_i:j\geq D_i
\right\}.
\end{equation}
Values in $\mis_i^{\pre}$ are imputed from the model for the assigned arm, whereas those in $\mis_i^{\post}$ follow the reference-based rule. This definition allows intermittent missingness before the trigger. When $D_i=J+1$, the set $\mis_i^{\post}$ is empty, and all unrecorded visits are handled under the model for the assigned arm.

Subvectors indexed by visit sets are always ordered by increasing visit number. The retained count vector available for model fitting and conditioning is
\begin{equation}
Y_{i,\obs_i}
=
\left(
Y_{ij}:j\in\obs_i
\right)^\top,
\end{equation}
and the target of imputation is
\begin{equation}
Y_{i,\mis_i}
=
\left(
Y_{ij}:j\in\mis_i
\right)^\top.
\end{equation}
The corresponding realized vectors are denoted by $y_{i,\obs_i}$ and $y_{i,\mis_i}$.

Let $r$ index a prespecified imputation rule, and let $p_r$ denote probabilities under the joint model defined by that rule. For the realized visit sets and trigger index, the conditional distribution used to impute the missing counts is
\begin{equation}
p_r\left(
y_{i,\mis_i}
\middle|
y_{i,\obs_i},
A_i,
X_i,
e_i,
D_i
\right).
\end{equation}

\subsection{Negative binomial marginal model}

We begin with an NB margin at each visit. These margins define the count distributions on which the reference-based assumptions are imposed.

Here and throughout, ``marginal'' refers to the visit-specific count distribution conditional on treatment, baseline covariates, and exposure, but not on the participant's retained counts. We use $a \in \left\{0, 1\right\}$ for a generic treatment value, with $a = 0$ for the reference arm and $a = 1$ for the active arm. Both $X_i$ and $e_{ij}$ are treated as fixed conditioning quantities.

For participant $i$, visit $j$, and treatment arm $a \in \left\{0, 1\right\}$, let
\begin{equation}
\mu_{ij}^{a}
=
\E\left(Y_{ij} \middle| A_i = a, X_i, e_{ij}\right)
\end{equation}
denote the marginal mean count under arm $a$, conditional on $X_i$ and $e_{ij}$. Write
\begin{equation}
\mu_{ij}^{a}
=
e_{ij}\lambda_{ij}^{a},
\end{equation}
where $\lambda_{ij}^{a}$ is the marginal rate under arm $a$. We assume
\begin{equation}
\left. Y_{ij} \middle| A_i = a, X_i, e_{ij} \right.
\sim
\NB\left(\mu_{ij}^{a}, \alpha_j\right),
\end{equation}
where $\NB\left(\cdot,\cdot\right)$ denotes an NB distribution parameterized by its mean in the first argument and its overdispersion parameter in the second argument, such that
\begin{equation}
\Var\left(Y_{ij} \middle| A_i = a, X_i, e_{ij}\right)
=
\mu_{ij}^{a}
+
\alpha_j\left(\mu_{ij}^{a}\right)^2.
\end{equation}
Here, $\alpha_j > 0$ is the visit-specific overdispersion parameter.

We model the marginal rate on the log scale as
\begin{equation}
\log\left(\lambda_{ij}^{a}\right)
=
\eta_j
+
\tau_j a
+
X_i^\top\beta_j,
\end{equation}
or equivalently,
\begin{equation}
\log\left(\mu_{ij}^{a}\right)
=
\log\left(e_{ij}\right)
+
\eta_j
+
\tau_j a
+
X_i^\top\beta_j.
\end{equation}
Here, $\eta_j$ is the log rate at visit $j$ in the reference arm when $X_i = 0$, $\tau_j$ is the log rate ratio comparing the active and reference arms at that visit, and $\beta_j \in \R^p$ is the column vector of covariate effects on the log scale. Exposure is fixed in the imputation model, and $\log\left(e_{ij}\right)$ enters as an offset.

Let
\begin{equation}
\pi_{ij}^{a}\left(y\right)
=
\Pr\left(Y_{ij} = y \middle| A_i = a, X_i, e_{ij}\right),
\quad
y \in \Nzero,
\end{equation}
denote the probability mass function of this NB distribution. The corresponding cumulative distribution function (CDF) is
\begin{equation}
F_{ij}^{a}\left(y\right)
=
\Pr\left(Y_{ij} \leq y \middle| A_i = a, X_i, e_{ij}\right)
=
\sum_{k = 0}^{y}
\pi_{ij}^{a}\left(k\right),
\quad
y \in \Nzero,
\end{equation}
with the convention
\begin{equation}
F_{ij}^{a}\left(-1\right)
=
0.
\end{equation}
Thus,
\begin{equation}
\pi_{ij}^{a}\left(y\right)
=
F_{ij}^{a}\left(y\right)
-
F_{ij}^{a}\left(y - 1\right).
\end{equation}
The functions $F_{ij}^{0}$ and $F_{ij}^{1}$ define the marginal count distributions in the reference and active arms. To keep the notation concise, we suppress the dependence of $\mu_{ij}^{a}$, $\pi_{ij}^{a}$, and $F_{ij}^{a}$ on $X_i$ and $e_{ij}$ below.

\subsection{Gaussian copula dependence model}

A Gaussian copula provides a joint distribution for the longitudinal count vector while preserving its NB margins. Throughout, $\Normal\left(\cdot,\cdot\right)$ denotes a normal distribution parameterized by its mean or mean vector in the first argument and its variance or covariance matrix in the second argument.

For participant $i$, let
\begin{equation}
Z_i =
\left(
Z_{i1}, \ldots, Z_{iJ}
\right)^\top
\end{equation}
be a latent Gaussian vector with
\begin{equation}
Z_i \sim \Normal\left(0, R\left(\psi\right)\right),
\end{equation}
where $R\left(\psi\right)$ is a $J \times J$ latent Gaussian correlation matrix and $\psi$ is the dependence parameter vector on its natural scale. Its parameter space is restricted to values for which $R\left(\psi\right)$ is a valid correlation matrix. The latent vectors are assumed independent across participants, and $R\left(\psi\right)$ is common across participants and treatment arms. An unconstrained parameterization for computation is introduced in \autoref{subsec:pit-augmentation}. We write $R$ when the dependence parameter is fixed. Each component of $Z_i$ corresponds to a scheduled visit, and the copula uniforms are
\begin{equation}
U_{ij}
=
\Phi\left(Z_{ij}\right),
\end{equation}
where $\Phi$ is the standard normal CDF.

For a participant assigned to arm $A_i = a$, the count at visit $j$ is obtained from the NB quantile function:
\begin{equation}
Y_{ij}
=
\inf
\left\{
y \in \Nzero:
F_{ij}^{a}\left(y\right) \geq U_{ij}
\right\}.
\end{equation}
This gives
\begin{equation}
Y_{ij}
\sim
\NB\left(\mu_{ij}^{a}, \alpha_j\right)
\end{equation}
marginally, with dependence among repeated counts induced by $R\left(\psi\right)$.

With discrete margins, the copula in Sklar's representation is uniquely determined only on the Cartesian product of the ranges of the marginal CDFs; its extension to the remainder of $\left[0,1\right]^J$ is not unique \citep{genest2007primer}. We therefore interpret the Gaussian copula as a parametric latent variable working model that induces a valid joint distribution for the scheduled counts. The entries of $R\left(\psi\right)$ are correlations on the latent Gaussian scale. Dependence among the observed counts is determined jointly by these correlations and the NB margins.

For a count retained for conditioning, define the marginal distribution according to the assigned arm as
\begin{equation}
F_{ij}\left(y\right)
=
F_{ij}^{A_i}\left(y\right),
\end{equation}
evaluated at $X_i$ and $e_{ij}$. Its probability mass is
\begin{equation}
\pi_{ij}\left(y\right)
=
F_{ij}\left(y\right)
-
F_{ij}\left(y - 1\right).
\end{equation}
This margin is used in the likelihood for the retained counts and in the randomized PIT augmentation. The margins used under each rule from the trigger onward are defined separately below.

Because the margins are discrete, $Y_{ij} = y$ does not identify a unique latent uniform or Gaussian value. Instead,
\begin{equation}
F_{ij}\left(y - 1\right)
<
U_{ij}
\leq
F_{ij}\left(y\right),
\end{equation}
or, equivalently,
\begin{equation}
\Phi^{-1}
\left\{
F_{ij}\left(y - 1\right)
\right\}
<
Z_{ij}
\leq
\Phi^{-1}
\left\{
F_{ij}\left(y\right)
\right\}.
\end{equation}
Only counts indexed by $\obs_i$ contribute to the retained count likelihood, and these visits need not be consecutive. For visit sets $\mathcal{A}, \mathcal{B} \subseteq \left\{1, \ldots, J\right\}$, let $R_{\mathcal{A}\mathcal{B}}\left(\psi\right)$ be the submatrix of $R\left(\psi\right)$ with rows indexed by $\mathcal{A}$ and columns by $\mathcal{B}$. For any visit set $\mathcal{A} \subseteq \left\{1,\ldots,J\right\}$, let $Z_{i,\mathcal{A}}$ denote the corresponding subvector of $Z_i$, and let $z_{i,\mathcal{A}}$ denote a generic value of that subvector. In particular, $R_{\obs_i\obs_i}\left(\psi\right)$ is the correlation matrix for the visits in $\obs_i$. We use $\left|\obs_i\right|$ for the number of retained visits, $I_{\left|\obs_i\right|}$ for the identity matrix of that dimension, and $\phi_{\left|\obs_i\right|}\left(\cdot;0,R_{\obs_i\obs_i}\left(\psi\right)\right)$ for the corresponding multivariate normal density with mean zero.

The retained counts indexed by $\obs_i$ define the following rectangle on the latent Gaussian scale:
\begin{equation}
\mathcal{B}_{i,\obs_i}
=
\left\{
z_{i,\obs_i}:
\Phi^{-1}
\left[
F_{ij}\left(y_{ij} - 1\right)
\right]
<
z_{ij}
\leq
\Phi^{-1}
\left[
F_{ij}\left(y_{ij}\right)
\right],
\ j \in \obs_i
\right\}.
\end{equation}
Let $\theta$ denote the full vector of marginal and copula parameters on their natural scales. Treating $\obs_i$ as fixed, the exact likelihood contribution for the retained counts is the multivariate normal rectangle probability
\begin{equation}
\Pr\left(Y_{i,\obs_i} = y_{i,\obs_i} \middle| A_i, X_i, e_i, \theta\right)
=
\int_{\mathcal{B}_{i,\obs_i}}
\phi_{\left|\obs_i\right|}
\left(
z_{i,\obs_i}; 0, R_{\obs_i\obs_i}\left(\psi\right)
\right)
d z_{i,\obs_i}.
\end{equation}
When $\obs_i=\emptyset$, we define the corresponding zero-dimensional Gaussian probability to be one.

Although this integral clarifies the model, evaluating it directly can be cumbersome when more than a few visits are observed. The randomized PIT augmentation in the next subsection avoids direct calculation of these rectangle probabilities.

\subsection{Randomized probability integral transform augmentation}
\label{subsec:pit-augmentation}

The interval for $Z_{ij}$ shows why an observed count cannot be replaced by a single deterministic normal score. Our randomized PIT augmentation applies standard data augmentation arguments for copula models with discrete margins \citep{smith2012estimation} to the NB Gaussian copula model.

Specifically, write this parameter vector as
\begin{equation}
\theta
=
\left(
\eta_1,\ldots,\eta_J,
\tau_1,\ldots,\tau_J,
\beta_1^\top,\ldots,\beta_J^\top,
\alpha_1,\ldots,\alpha_J,
\psi^\top
\right)^\top.
\end{equation}
For posterior computation, define the unconstrained parameter vector
\begin{equation}
\vartheta
=
\left(
\eta_1,\ldots,\eta_J,
\tau_1,\ldots,\tau_J,
\beta_1^\top,\ldots,\beta_J^\top,
\xi_1,\ldots,\xi_J,
\psi_{\mathrm{raw}}^\top
\right)^\top,
\end{equation}
where
\begin{equation}
\xi_j
=
\log\left(\alpha_j\right),
\qquad
\theta
=
h\left(\vartheta\right).
\end{equation}
The transformation $h$ maps $\xi_j$ to $\alpha_j=\exp\left(\xi_j\right)$ and $\psi_{\mathrm{raw}}$ to a valid dependence parameter vector $\psi$ on the natural scale. The transformation for $\psi_{\mathrm{raw}}$ depends on the form of $R\left(\psi\right)$. For the first-order autoregressive (AR(1)) structure used in the simulations and application, it is given in \autoref{subsec:estimation-implementation}. If the marginal model is more parsimonious, for example because covariate effects are common across visits, $\vartheta$ and $\theta$ contain the corresponding lower-dimensional parameterization. We write $F_{ij}\left(\cdot;\theta\right)$ and $\pi_{ij}\left(\cdot;\theta\right)$ for the CDF and probability mass under the assigned arm, evaluated at $\theta$.

For computation, introduce one auxiliary PIT variable for every retained count. Under the joint base measure, these variables are mutually independent, with
\begin{equation}
V_{ij}
\sim
\operatorname{Uniform}\left(0,1\right),
\qquad
j \in \obs_i,
\quad
i=1,\ldots,n,
\end{equation}
where the two arguments of $\operatorname{Uniform}\left(\cdot,\cdot\right)$ denote the lower and upper bounds, respectively.

Define
\begin{equation}
Q_{ij}\left(\theta,V_{ij}\right)
=
F_{ij}\left(y_{ij} - 1;\theta\right)
+
V_{ij}
\pi_{ij}\left(y_{ij};\theta\right).
\end{equation}
The corresponding latent Gaussian value is
\begin{equation}
Z_{ij}^{*}\left(\theta,V_{ij}\right)
=
\Phi^{-1}
\left[
Q_{ij}\left(\theta,V_{ij}\right)
\right]
\end{equation}
and lies within the latent Gaussian interval implied by the retained count.

For participant $i$, collect the PIT variables as
\begin{equation}
V_{i,\obs_i}
=
\left(
V_{ij}:j\in\obs_i
\right)^\top,
\end{equation}
and let
\begin{equation}
V_{\obs}
=
\left\{
V_{ij}:j\in\obs_i,\ i=1,\ldots,n
\right\}
\end{equation}
denote the full collection across participants. Define
\begin{equation}
Q_{i,\obs_i}
\left(
\theta,V_{i,\obs_i}
\right)
=
\left(
Q_{ij}\left(\theta,V_{ij}\right):
j\in\obs_i
\right)^\top,
\end{equation}
ordered by visit, and let
\begin{equation}
Z_{i,\obs_i}^{*}
\left(
\theta,V_{i,\obs_i}
\right)
=
\Phi^{-1}
\left[
Q_{i,\obs_i}
\left(
\theta,V_{i,\obs_i}
\right)
\right],
\end{equation}
where $\Phi^{-1}$ is applied componentwise.

For compactness, write
\begin{equation}
q_{i,\obs_i}
=
Q_{i,\obs_i}
\left(
\theta,V_{i,\obs_i}
\right),
\qquad
z_{i,\obs_i}^{*}
=
Z_{i,\obs_i}^{*}
\left(
\theta,V_{i,\obs_i}
\right).
\end{equation}
Let $c_{R_{\obs_i\obs_i}\left(\psi\right)}$ denote the Gaussian copula density associated with $R_{\obs_i\obs_i}\left(\psi\right)$. For the retained visits,
\begin{equation}
c_{R_{\obs_i\obs_i}\left(\psi\right)}
\left(
q_{i,\obs_i}
\right)
=
\left|
R_{\obs_i\obs_i}\left(\psi\right)
\right|^{-1/2}
\exp
\left[
-\frac{1}{2}
\left(z_{i,\obs_i}^{*}\right)^\top
\left(
\left\{R_{\obs_i\obs_i}\left(\psi\right)\right\}^{-1}
-
I_{\left|\obs_i\right|}
\right)
z_{i,\obs_i}^{*}
\right].
\end{equation}
The contribution to the augmented likelihood is
\begin{equation}
L_i^{\aug}\left(\theta,V_{i,\obs_i}\right)
=
c_{R_{\obs_i\obs_i}\left(\psi\right)}
\left(
q_{i,\obs_i}
\right)
\prod_{j \in \obs_i}
\pi_{ij}
\left(
y_{ij};\theta
\right).
\end{equation}
When $\obs_i=\emptyset$, the copula density and empty product are defined to be one, so that
\begin{equation}
L_i^{\aug}\left(\theta,V_{i,\obs_i}\right)
=
1.
\end{equation}
The PIT variables have a uniform base measure, but the augmented likelihood determines their conditional distribution, which is generally weighted by the Gaussian copula density. Integrating them out recovers the exact rectangle probability for the retained counts. \appref{app:pit-augmentation} gives the derivation for the NB Gaussian copula model. The PIT variables therefore represent uncertainty about the latent position of each retained count within its probability interval. They must be integrated over, either analytically or by sampling, rather than drawn once and treated as fixed transformed data.

\subsection{Reference-based imputation rules}

The reference-based assumptions specify the NB margins used to complete missing values from the trigger onward for participants in the active arm. They define the imputation component of the sensitivity analysis, not the full estimand, which also depends on the target population, treatment condition, endpoint, strategy for intercurrent events, and treatment effect summary. At each visit $j \in \mis_i$, rule $r$ specifies a marginal mean $\mu_{ij}^{r}$ and hence an NB distribution
\begin{equation}
Y_{ij}^{r}
\sim
\NB\left(\mu_{ij}^{r}, \alpha_j\right).
\end{equation}
The rule acts through the marginal mean used for completion. The visit-specific overdispersion parameter $\alpha_j$ and the fitted Gaussian copula correlation structure are retained across rules. Although $\alpha_j$ remains fixed, the NB variance changes with the rule-specific mean.

For participant $i$ at visit $j$, the marginal mean under the assigned arm is
\begin{equation}
\mu_{ij}^{\asgn}
=
\mu_{ij}^{A_i}
=
e_{ij}\lambda_{ij}^{A_i},
\end{equation}
and the mean under the reference arm is
\begin{equation}
\mu_{ij}^{\refarm}
=
\mu_{ij}^{0}
=
e_{ij}\lambda_{ij}^{0}.
\end{equation}
Both means use the same exposure $e_{ij}$ for that participant and visit. The rule changes the rate used to complete the count, not its exposure.

Values requiring imputation before the trigger follow the distribution for the assigned arm:
\begin{equation}
\mu_{ij}^{r}
=
\mu_{ij}^{\asgn},
\quad
j \in \mis_i^{\pre}.
\end{equation}
This means that intermittent missing values before the trigger are completed under the model for the assigned arm.

For values from the trigger onward, define a controlled family of reference-based rules indexed by the prespecified vector
\begin{equation}
\omega
=
\left(
\omega_1,\ldots,\omega_J
\right)^\top,
\qquad
\omega_j \in \left[0,1\right].
\end{equation}
Within this family, rule $r$ corresponds to a specified choice of $\omega$, and we write $\mu_{ij}^{\omega}$ when displaying the rule-specific mean explicitly. For participants in the active arm,
\begin{equation}
\log\left(\mu_{ij}^{\omega}\right)
=
\left(1 - \omega_j\right)
\log\left(\mu_{ij}^{\asgn}\right)
+
\omega_j
\log\left(\mu_{ij}^{\refarm}\right),
\quad
A_i = 1,\ j \in \mis_i^{\post}.
\end{equation}
Equivalently,
\begin{equation}
\mu_{ij}^{\omega}
=
\left(\mu_{ij}^{\asgn}\right)^{1 - \omega_j}
\left(\mu_{ij}^{\refarm}\right)^{\omega_j}.
\end{equation}
Using the log rate model, this can also be written as
\begin{equation}
\log\left(\mu_{ij}^{\omega}\right)
=
\log\left(e_{ij}\right)
+
\eta_j
+
X_i^\top\beta_j
+
\left(1 - \omega_j\right)\tau_j,
\quad
A_i = 1,\ j \in \mis_i^{\post}.
\end{equation}
The value of $\omega_j$ controls the shift from the marginal mean under the assigned arm toward that under the reference arm. On the log rate scale, $1-\omega_j$ is interpreted here as the proportion of the treatment effect retained. Because $\omega_j$ encodes an assumption about unobserved outcomes after the trigger, it is specified as a sensitivity parameter rather than estimated. We use $\omega_j=0.5$ as an illustrative midpoint; other values can be prespecified on clinical grounds or explored over a grid \citep{cro2020sensitivity,white2020causal}.

Assigned arm continuation sets
\begin{equation}
\omega_j = 0,
\qquad
j \in \mis_i^{\post}.
\end{equation}
For participants in the active arm, values in $\mis_i^{\post}$ therefore retain the NB margins under the assigned arm.

J2R sets
\begin{equation}
\omega_j = 1,
\qquad
j \in \mis_i^{\post}.
\end{equation}
For participants in the active arm, values in $\mis_i^{\post}$ therefore use the NB margins under the reference arm. Values with $0 < \omega_j < 1$ define controlled rules between assigned arm continuation and J2R.

For participants in the reference arm, the assigned arm is the reference arm, so
\begin{equation}
\mu_{ij}^{r}
=
\mu_{ij}^{0},
\quad
A_i = 0,\ j \in \mis_i,
\end{equation}
and all of the rules coincide.

The CDF under rule $r$ is denoted by
\begin{equation}
F_{ij}^{r}\left(y\right)
=
\Pr\left(Y_{ij}^{r} \leq y \middle| A_i, X_i, e_{ij}\right),
\end{equation}
where the NB distribution has mean $\mu_{ij}^{r}$, based on exposure $e_{ij}$, and overdispersion $\alpha_j$. Its probability mass is
\begin{equation}
\pi_{ij}^{r}\left(y\right)
=
F_{ij}^{r}\left(y\right)
-
F_{ij}^{r}\left(y - 1\right).
\end{equation}
The reference-based rule specifies the NB margin for each value requiring completion. Combined with the fitted latent Gaussian dependence structure, these margins determine the joint conditional distribution of the counts in $\mis_i$ given those in $\obs_i$. Participants with the same treatment and covariates can consequently have different imputation distributions when their retained count profiles differ.

\begin{figure}[!t]
\centering
\includegraphics[width=0.90\textwidth]{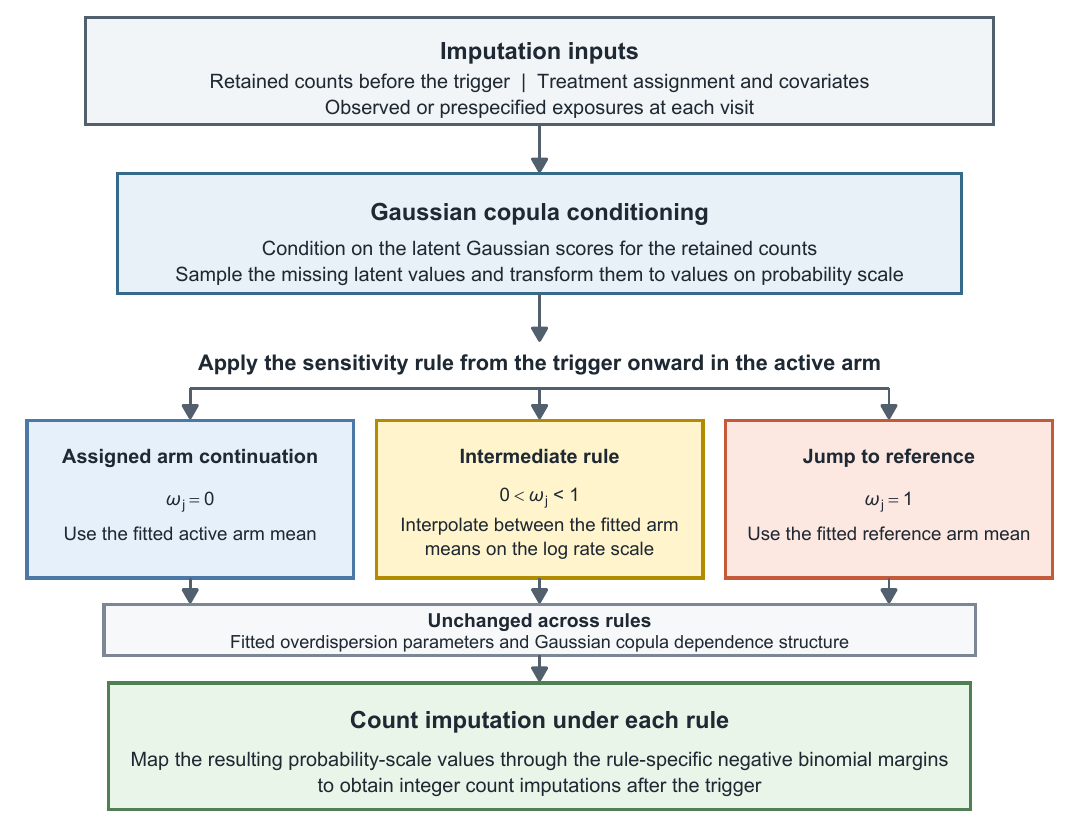}
\caption{Schematic of the proposed reference-based imputation procedure. Through the fitted copula, retained counts condition the distribution of the missing latent Gaussian values. For values requiring completion in the active arm from the trigger onward, the sensitivity rule changes the NB marginal mean, while the fitted overdispersion parameters and latent Gaussian correlation structure remain fixed across rules. The resulting values on the probability scale are mapped through the rule-specific NB margins to generate integer counts. All rules coincide in the reference arm. The schematic shows monotone missingness; any intermittently missing values before the trigger are completed under the margin for the assigned arm.}
\label{fig:rbmi-imputation-schematic}
\end{figure}

The full procedure is summarized in
\autoref{fig:rbmi-imputation-schematic}.

\subsection{Conditional imputation algorithm}
\label{subsec:conditional-imputation}

The joint NB Gaussian copula model induces the conditional distribution to be imputed. We treat the realized trigger index $D_i$, which determines $\obs_i$ and $\mis_i$, as fixed. \appref{app:conditional-distribution} expresses this distribution in terms of latent rectangles and shows that its denominator, the probability of the retained counts, is the same under every reference-based rule. We sample from the conditional distribution through the latent Gaussian model rather than evaluate the discrete probabilities directly.

The posterior sampler targets the joint augmented posterior of the computational parameter vector $\vartheta$ and all auxiliary PIT variables
\begin{equation}
V_{\obs}
=
\left\{
V_{ij}:
j \in \obs_i,\ i = 1,\ldots,n
\right\},
\end{equation}
with one $V_{ij}$ for each retained count $y_{ij}$. Let $S$ be the total number of post-warmup draws retained across all chains. The $s$th joint draw is
\begin{equation}
\left(
\vartheta^{\left(s\right)},
V_{\obs}^{\left(s\right)}
\right),
\quad
s = 1,\ldots,S,
\end{equation}
where $V_{\obs}^{\left(s\right)}$ contains the PIT variables from the same posterior draw as $\vartheta^{\left(s\right)}$. In particular, $V_{ij}^{\left(s\right)}$ gives the sampled relative position within the PIT interval associated with retained count $y_{ij}$ at draw $s$. The corresponding parameter vector on the natural scale is
\begin{equation}
\theta^{\left(s\right)}
=
h\left(\vartheta^{\left(s\right)}\right).
\end{equation}
\autoref{alg:conditional-rbmi} describes the procedure for participant $i$, posterior draw $s$, and rule $r$. It is repeated for every participant and each posterior draw required for the analysis. Posterior predictive summaries can use all $S$ draws. For conventional multiple imputation, one may instead select $M$ joint posterior draws and generate one completed dataset from each.

\begin{algorithm}[!htbp]
\caption{Conditional reference-based imputation for participant $i$ under rule $r$}
\label{alg:conditional-rbmi}
\centering
\fbox{
\begin{minipage}{0.97\textwidth}
\footnotesize
\begin{enumerate}[leftmargin=1.75em, itemsep=2pt]
\item Take the joint posterior draw
\begin{equation}
\left(
\vartheta^{\left(s\right)},
V_{\obs}^{\left(s\right)}
\right),
\end{equation}
set $\theta^{\left(s\right)}=h\left(\vartheta^{\left(s\right)}\right)$ and $R^{\left(s\right)}=R\left(\psi^{\left(s\right)}\right)$, and extract the subset $V_{i,\obs_i}^{\left(s\right)}$ for participant $i$.

\item If $\mis_i=\emptyset$, stop because participant $i$ requires no imputation. If $\obs_i=\emptyset$, sample
\begin{equation}
Z_{i,\mis_i}^{\misval,\left(s\right)}
\sim
\Normal\left(
0,
R_{\mis_i\mis_i}^{\left(s\right)}
\right)
\end{equation}
and continue with Step~6. Otherwise, proceed with the conditioning steps below.

\item For each conditioning visit $j \in \obs_i$, use the randomized PIT augmentation to calculate the latent Gaussian score $Z_{ij}^{*,\left(s\right)}$.

\item Select the blocks of $R^{\left(s\right)}$ indexed by $\obs_i$ and $\mis_i$.

\item Draw $Z_{i,\mis_i}^{\misval,\left(s\right)}$ from its conditional normal distribution given $Z_{i,\obs_i}^{*,\left(s\right)}$.

\item Transform the missing latent Gaussian values to values on the probability scale, denoted by $U_{ij}^{\misval,\left(s\right)}$, $j \in \mis_i$.

\item Map each $U_{ij}^{\misval,\left(s\right)}$ through the NB distribution $F_{ij}^{r,\left(s\right)}$ for rule $r$, evaluated using $e_{ij}$, to obtain $Y_{ij}^{\misval,\left(s,r\right)}$.

\item Form the completed vector $Y_i^{\comp,\left(s,r\right)}$ by retaining $Y_{ij}$ for $j \in \obs_i$ and replacing $Y_{ij}$ with $Y_{ij}^{\misval,\left(s,r\right)}$ for $j \in \mis_i$.
\end{enumerate}
\end{minipage}
}
\end{algorithm}

For $j \in \obs_i$, evaluate the CDF under the assigned arm at draw $s$:
\begin{equation}
F_{ij}^{\left(s\right)}\left(y\right)
=
F_{ij}^{A_i}\left(
y;
\theta^{\left(s\right)}
\right).
\end{equation}
The randomized PIT augmentation gives the following latent Gaussian score for retained count $y_{ij}$:
\begin{equation}
Z_{ij}^{*,\left(s\right)}
=
\Phi^{-1}
\left[
F_{ij}^{\left(s\right)}\left(y_{ij}-1\right)
+
V_{ij}^{\left(s\right)}
\left\{
F_{ij}^{\left(s\right)}\left(y_{ij}\right)
-
F_{ij}^{\left(s\right)}\left(y_{ij}-1\right)
\right\}
\right].
\end{equation}
Here, $V_{ij}^{\left(s\right)}$ is the auxiliary PIT variable from posterior draw $s$, so $Z_{ij}^{*,\left(s\right)}$ is a sampled position within the latent Gaussian interval implied by $y_{ij}$. Let
\begin{equation}
Z_{i,\obs_i}^{*,\left(s\right)}
=
\left(
Z_{ij}^{*,\left(s\right)}:
j \in \obs_i
\right)^\top
\end{equation}
collect these latent Gaussian scores for the retained counts.

For a participant with $\obs_i\neq\emptyset$ and $\mis_i\neq\emptyset$, arrange the latent Gaussian components with visits in $\obs_i$ followed by those in $\mis_i$. Applying the same permutation to the rows and columns of the correlation matrix gives the following block form at draw $s$:
\begin{equation}
\begin{pmatrix}
R_{\obs_i\obs_i}^{\left(s\right)}
&
R_{\obs_i\mis_i}^{\left(s\right)}
\\
R_{\mis_i\obs_i}^{\left(s\right)}
&
R_{\mis_i\mis_i}^{\left(s\right)}
\end{pmatrix},
\end{equation}
where the submatrices are obtained from
\begin{equation}
R^{\left(s\right)}
=
R\left(\psi^{\left(s\right)}\right)
\end{equation}
by selecting the rows and columns shown in the subscripts. The sets $\obs_i$ and $\mis_i$ need not contain consecutive visits.

Conditional on the latent Gaussian scores for the retained counts, the missing latent Gaussian vector follows
\begin{equation}
\left.
Z_{i,\mis_i}^{\misval,\left(s\right)}
\middle|
Z_{i,\obs_i}^{*,\left(s\right)}
\right.
\sim
\Normal\left(
m_{i,\left.\mis_i \middle| \obs_i\right.}^{\left(s\right)},
\Sigma_{i,\left.\mis_i \middle| \obs_i\right.}^{\left(s\right)}
\right),
\end{equation}
where
\begin{equation}
m_{i,\left.\mis_i \middle| \obs_i\right.}^{\left(s\right)}
=
R_{\mis_i\obs_i}^{\left(s\right)}
\left(
R_{\obs_i\obs_i}^{\left(s\right)}
\right)^{-1}
Z_{i,\obs_i}^{*,\left(s\right)}
\end{equation}
and
\begin{equation}
\Sigma_{i,\left.\mis_i \middle| \obs_i\right.}^{\left(s\right)}
=
R_{\mis_i\mis_i}^{\left(s\right)}
-
R_{\mis_i\obs_i}^{\left(s\right)}
\left(
R_{\obs_i\obs_i}^{\left(s\right)}
\right)^{-1}
R_{\obs_i\mis_i}^{\left(s\right)}.
\end{equation}
When $\obs_i=\emptyset$, no latent Gaussian scores are available for conditioning and $\mis_i=\left\{1,\ldots,J\right\}$. The missing vector is then sampled from its unconditional distribution, as in Step~2 of \autoref{alg:conditional-rbmi}.

The sampled latent Gaussian values are transformed to values on the probability scale:
\begin{equation}
U_{ij}^{\misval,\left(s\right)}
=
\Phi\left(
Z_{ij}^{\misval,\left(s\right)}
\right),
\quad
j \in \mis_i.
\end{equation}
When $\obs_i \neq \emptyset$, these values are generally not uniformly distributed conditional on the retained latent Gaussian scores. When $\obs_i = \emptyset$, each has a marginal $\operatorname{Uniform}\left(0,1\right)$ distribution, although the components need not be mutually independent.

For each $j \in \mis_i$, evaluate the CDF for rule $r$ at draw $s$:
\begin{equation}
F_{ij}^{r,\left(s\right)}\left(y\right)
=
F_{ij}^{r}\left(
y;
\theta^{\left(s\right)}
\right).
\end{equation}
The value $U_{ij}^{\misval,\left(s\right)}$ is then mapped to an integer count using the corresponding NB quantile function:
\begin{equation}
Y_{ij}^{\misval,\left(s,r\right)}
=
\inf
\left\{
y \in \Nzero:
F_{ij}^{r,\left(s\right)}\left(y\right)
\geq
U_{ij}^{\misval,\left(s\right)}
\right\},
\quad
j \in \mis_i.
\end{equation}
The distribution $F_{ij}^{r,\left(s\right)}$ has mean $\mu_{ij}^{r,\left(s\right)}$, which incorporates exposure $e_{ij}$, and visit-specific overdispersion $\alpha_j^{\left(s\right)}$.

For visits in $\mis_i^{\pre}$, $F_{ij}^{r,\left(s\right)}$ is the distribution under the assigned arm. For visits in $\mis_i^{\post}$ among participants in the active arm, the chosen reference-based rule determines this distribution. The distributions under the assigned and reference arms coincide for participants in the reference arm.

Several rules can be compared using the same joint posterior draw and sampled missing latent Gaussian vector, and therefore the same missing $U$ values. For visits in $\mis_i^{\post}$ among participants in the active arm, only the NB CDF used to map a given $U$ value to a count differs between rules. Reusing common random numbers reduces Monte Carlo variation in paired comparisons without changing the validity of the completed outcomes under any rule. Each completed dataset remains a draw from the conditional imputation distribution for that rule.

The completed count vector for participant $i$ under rule $r$ is
\begin{equation}
Y_i^{\comp,\left(s,r\right)}
=
\left(
Y_{ij}^{\comp,\left(s,r\right)}:
j = 1,\ldots,J
\right)^\top,
\end{equation}
where
\begin{equation}
Y_{ij}^{\comp,\left(s,r\right)}
=
\begin{cases}
Y_{ij}, & j \in \obs_i, \\
Y_{ij}^{\misval,\left(s,r\right)}, & j \in \mis_i.
\end{cases}
\end{equation}
Applying the procedure to all $S$ post-warmup draws yields samples from the posterior predictive distribution of the completed count vectors. For conventional multiple imputation, applying it to $M$ prespecified draws produces $M$ completed datasets under each rule. Estimates from these datasets are combined using the Rubin pooling procedure in \autoref{subsec:rubin-pooling}.

\subsection{Estimation and implementation}
\label{subsec:estimation-implementation}

The imputation model is fitted to the counts retained for conditioning:
\begin{equation}
\mathcal{D}_{\obs}
=
\left\{
Y_{i,\obs_i}, A_i, X_i, e_i:
i = 1,\ldots,n
\right\}.
\end{equation}
Values in $\mis_i$ are treated as missing during model fitting. These include both unrecorded values and, where applicable, recorded values from the trigger onward that are excluded from the conditioning set. The reference-based rule is not used to estimate the model. It enters only when the posterior predictive distribution is used to complete the data.

The likelihood is based on $Y_{i,\obs_i}$, treating the realized conditioning set $\obs_i$ and, when relevant, the intercurrent event timing that determines it as fixed. Neither the recording indicators nor the intercurrent event process is modeled jointly with the outcomes. As with standard likelihood-based longitudinal imputation, the fitted model should therefore be viewed as a working imputation model if missingness or the intercurrent event depends on unobserved outcomes beyond the counts and covariates used for conditioning.

Posterior computation uses the unconstrained parameter vector $\vartheta$ defined in \autoref{subsec:pit-augmentation}. The quantities $F_{ij}$, $\pi_{ij}$, and $R$ entering the likelihood are evaluated on their natural scales at $\theta=h\left(\vartheta\right)$.

Unless stated otherwise, independent priors are assigned directly to the components of $\vartheta$. The marginal log rate parameters have diffuse normal priors,
\begin{equation}
\eta_j \sim \Normal\left(0, 1000\right),
\quad
\tau_j \sim \Normal\left(0, 1000\right),
\quad
\beta_{jk} \sim \Normal\left(0, 1000\right),
\end{equation}
for $j=1,\ldots,J$ and $k=1,\ldots,p$. The unconstrained overdispersion coordinates have priors
\begin{equation}
\xi_j
\sim
\Normal\left(0, 4\right),
\qquad
\alpha_j
=
\exp\left(\xi_j\right).
\end{equation}
For the AR(1) structure used in the simulations and application, dependence is indexed by scheduled visit order rather than elapsed study week. Let $\operatorname{pos}\left(j\right)\in\{1,\ldots,J\}$ denote the ordinal position of visit $j$. The latent correlation matrix has entries
\begin{equation}
R_{jk}\left(\rho\right)
=
\rho^{
\left|
\operatorname{pos}\left(j\right)
-
\operatorname{pos}\left(k\right)
\right|
}.
\end{equation}
The dependence parameter on the natural scale is $\psi=\rho$, with unconstrained coordinate $\psi_{\mathrm{raw}}=\rho_{\mathrm{raw}}$. Its prior and transformation are
\begin{equation}
\rho_{\mathrm{raw}}
\sim
\Normal\left(0, 1\right),
\qquad
\rho
=
2\Phi\left(\rho_{\mathrm{raw}}\right)-1.
\end{equation}
Under this prior, $\Phi\left(\rho_{\mathrm{raw}}\right)\sim\mathrm{Uniform}\left(0,1\right)$, which induces a $\mathrm{Uniform}\left(-1,1\right)$ prior on $\rho$.

Using the augmented likelihood in \autoref{subsec:pit-augmentation}, the log posterior kernel on the computational scale is, up to an additive constant,
\begin{equation}
\ell_{\aug}\left(\vartheta,V_{\obs}\right)
=
\log p_{\vartheta}\left(\vartheta\right)
+
\sum_{i=1}^{n}
\log
L_i^{\aug}
\left(
h\left(\vartheta\right),
V_{i,\obs_i}
\right),
\end{equation}
where $p_{\vartheta}\left(\vartheta\right)$ is the joint prior density on the unconstrained computational coordinates.

This density is defined with respect to Lebesgue measure on $\vartheta$ and the uniform base measure for $V_{\obs}$. Because the priors are specified directly on $\vartheta$, $\ell_{\aug}$ requires no additional Jacobian for the transformation $h$; the transformation is used to evaluate the likelihood on the natural parameter scale. For fixed $\vartheta$, integrating over $V_{\obs}$ recovers the likelihood for the retained counts at $\theta=h\left(\vartheta\right)$. The PIT variables are therefore nuisance latent quantities sampled jointly with the computational parameters, not fixed transformations of the observed counts.

We target $\ell_{\aug}\left(\vartheta,V_{\obs}\right)$ with a custom Metropolis-within-Gibbs sampler. Each iteration has three types of updates:
\begin{enumerate}[leftmargin=1.75em,itemsep=2pt]
\item block updates of the unconstrained marginal coordinates, including the log-overdispersion coordinates $\xi_j$;
\item updates of the unconstrained copula dependence coordinate $\psi_{\mathrm{raw}}$;
\item separate updates of the PIT variables $V_{i,\obs_i}$ for each participant.
\end{enumerate}
All parameter proposals are made on the unconstrained scale. Let $b$ index a block of $\vartheta$, with $\vartheta_b$ denoting that block and $\vartheta_{-b}$ the remaining components. The proposal covariance $S_b^{\mathrm{prop}}$ is a sampler tuning quantity, not a model parameter. We use the random-walk proposal
\begin{equation}
\vartheta_b^{\dagger}
=
\vartheta_b
+
\varepsilon_b,
\quad
\varepsilon_b
\sim
\Normal\left(0,S_b^{\mathrm{prop}}\right).
\end{equation}
Let $\vartheta^{\dagger}=\left(\vartheta_b^{\dagger},\vartheta_{-b}\right)$ be the full proposed vector, with components outside block $b$ held fixed. Its acceptance probability is
\begin{equation}
a_b
=
\min
\left\{
1,
\exp
\left[
\ell_{\aug}
\left(
\vartheta^{\dagger},V_{\obs}
\right)
-
\ell_{\aug}
\left(
\vartheta,V_{\obs}
\right)
\right]
\right\}.
\end{equation}
When $\psi_{\mathrm{raw}}$ is updated in the AR(1) model, only the copula part of $\ell_{\aug}$ must be recalculated. Conditional on the current marginal parameters and PIT variables, the NB masses and PIT intervals do not change.

The PIT variables are updated separately for each participant using a random-walk proposal on the logit scale. Let $\sigma_V^2$ be the proposal variance used for this update. For participant $i$, define
\begin{equation}
H_{ij}
=
\log
\left(
\frac{V_{ij}}{1 - V_{ij}}
\right),
\quad
j \in \obs_i.
\end{equation}
A proposal is generated on the logit scale:
\begin{equation}
H_{i,\obs_i}^{\dagger}
=
H_{i,\obs_i}
+
\varepsilon_i,
\quad
\varepsilon_i
\sim
\Normal
\left(
0,
\sigma_V^2 I_{\left|\obs_i\right|}
\right),
\end{equation}
and transformed back to
\begin{equation}
V_{ij}^{\dagger}
=
\frac{
\exp\left(H_{ij}^{\dagger}\right)
}{
1 + \exp\left(H_{ij}^{\dagger}\right)
},
\quad
j \in \obs_i.
\end{equation}
Let $q_{i,\obs_i}^{\dagger}$ be the randomized PIT values induced by $V_{i,\obs_i}^{\dagger}$. The NB mass terms do not change in this update, so the acceptance probability depends only on the change in the copula density for participant $i$ and the Jacobian of the logit transformation:
\begin{equation}
a_{V_i}
=
\min
\left\{
1,
\exp
\left[
\Delta_i^{\mathrm{cop}}
+
\Delta_i^{\mathrm{jac}}
\right]
\right\},
\end{equation}
where
\begin{equation}
\Delta_i^{\mathrm{cop}}
=
\log
c_{R_{\obs_i\obs_i}\left(\psi\right)}
\left(
q_{i,\obs_i}^{\dagger}
\right)
-
\log
c_{R_{\obs_i\obs_i}\left(\psi\right)}
\left(
q_{i,\obs_i}
\right)
\end{equation}
and
\begin{equation}
\Delta_i^{\mathrm{jac}}
=
\sum_{j \in \obs_i}
\left[
\log V_{ij}^{\dagger}
+
\log\left(1 - V_{ij}^{\dagger}\right)
-
\log V_{ij}
-
\log\left(1 - V_{ij}\right)
\right].
\end{equation}
This update samples the latent location of each retained count within its probability interval. Integrating over the PIT variables recovers the exact likelihood for the retained counts.

During warmup, proposal scales are adapted using acceptance rates over short windows. They are fixed before posterior draws are retained. Independent chains are run in parallel when possible. For the monitored core parameters, convergence and Monte Carlo accuracy are assessed using potential scale reduction factors (PSRFs) and approximate effective sample sizes. Acceptance rates for the parameter blocks and for each participant's PIT update provide additional diagnostics.

Further implementation details are summarized in \appref{app:posterior-sampler}.

\section{Simulation studies}
\label{sec:simulation}

We conducted two targeted simulation studies. The first assessed recovery of the model parameters when every post-baseline outcome was observed. The second compared treatment effect estimates after imputation with estimates from the corresponding complete data before missingness was imposed. In this second study, outcomes after the trigger were generated under assigned arm continuation, the intermediate rule, or J2R. The first study used parameter values similar to those in the application. For the second, we increased the treatment effects and the amount of missingness to make differences between the imputation rules more visible.

\subsection{Model recovery with fully observed outcomes}
\label{subsec:simulation-model-recovery}

The first study examined parameter recovery under the NB Gaussian copula model and the performance of the posterior sampler when all scheduled post-baseline counts were observed. Each simulated trial included $n = 150$ participants randomized 1:1 to the reference or active arm, with counts recorded at Weeks~4, 8, and 12. Throughout both simulation studies, visit subscripts 4, 8, and 12 denote study weeks. A continuous baseline covariate was generated as
\begin{equation}
X_i \sim \Normal\left(0,1\right),
\end{equation}
and exposure was fixed at $e_{ij} = 3$ for all participants and visits. Counts were generated under the model described in \autoref{sec:methods}, using parameter values close to those estimated in the application. The exact values are given in \supptabref{supptab:simulation-dgp}. Both data generation and model fitting used a single covariate coefficient, $\beta_X$, common to the three post-baseline visits.

We fitted the imputation model to each complete simulated dataset using the sampler with randomized PIT augmentation, without applying a reference-based completion step. Performance measures were bias, root mean square error (RMSE), the empirical standard deviation (SD) of posterior means, the mean posterior SD, and coverage of 95\% credible intervals. This study assessed model and sampler performance without the additional effects of missingness and reference-based completion.

Of 301 attempted replicates, 300 met the convergence criteria after the prespecified fitting attempts and were included.

\autoref{tab:simulation-model-recovery} summarizes the results for the $R_1 = 300$ included trials. Bias was small for the marginal mean, overdispersion, and latent dependence parameters, and posterior SDs were generally close to the empirical SDs of the posterior means. Coverage of the 95\% credible intervals ranged from 0.94 to 0.96, except for $\tau_4$, which had coverage of 0.92. The model and sampler therefore performed well in this setting, although coverage for $\tau_4$ was below the nominal level. Results for every parameter appear in \supptabref{supptab:model-recovery-full}.

\begin{table}[!htbp]
\centering
\caption{Model recovery for the NB Gaussian copula model with fully observed outcomes.}
\label{tab:simulation-model-recovery}
\small
\begin{tabular}{@{}lcccc@{}}
\toprule
\textbf{Parameter group} & \textbf{No. par.} & \textbf{Max. abs. bias} & \textbf{RMSE range} & \textbf{Coverage range} \\
\midrule
Intercepts $\eta_j$ & 3 & 0.007 & 0.130--0.138 & 0.95--0.96 \\
Treatment effects $\tau_j$ & 3 & 0.022 & 0.184--0.204 & 0.92--0.95 \\
Covariate effect $\beta_X$ & 1 & 0.008 & 0.069 & 0.95 \\
Overdispersion $\alpha_j$ & 3 & 0.027 & 0.154--0.187 & 0.94--0.95 \\
Latent correlation $\rho$ & 1 & 0.008 & 0.051 & 0.94 \\
\bottomrule
\end{tabular}
\begin{flushleft}
\footnotesize
\justifying
No. par. = number of parameters in the group. Coverage is the empirical coverage of the 95\% posterior credible interval. Bias and RMSE are on the model parameter scale.
\end{flushleft}
\end{table}

\subsection{Agreement between treatment effect estimates from imputed and complete data}
\label{subsec:simulation-rbmi-performance}

The second study evaluated reference-based imputation with monotone missingness after baseline. The reference arm rates, covariate distribution, fixed exposure, overdispersion parameters, and latent Gaussian AR(1) structure were the same as in the first study. We strengthened the treatment effects at Weeks~8 and 12 so that differences between the rules would be clearly visible. The treatment rate ratios $\exp\left(\tau_j\right)$ were 0.909, 0.650, and 0.600 at Weeks~4, 8, and 12, respectively.

Missingness was generated sequentially from the outcome trajectory under assigned arm continuation. Let $G_{i8}=1$ indicate dropout beginning at Week~8 and, among participants with $G_{i8}=0$, let $G_{i12}=1$ indicate dropout beginning at Week~12. Treatment was coded as $A_i=0$ for the reference arm and $A_i=1$ for the active arm. Let $Y_{ij}^{\left(0\right)}$ be the count generated under assigned arm continuation, for which $\omega_{\mathrm{true}}=0$. In replicate $\ell$, the dropout probabilities were
\begin{align}
\operatorname{logit}
\Pr\left(G_{i8}=1 \middle| A_i,X_i,Y_{i4}^{\left(0\right)}\right)
&=
\kappa_{8\ell}
+
0.20A_i
+
0.25X_i
+
0.70
\log\left(
\frac{Y_{i4}^{\left(0\right)}+0.5}{3}
\right),
\\
\operatorname{logit}
\Pr\left(G_{i12}=1 \middle| G_{i8}=0,A_i,X_i,Y_{i8}^{\left(0\right)}\right)
&=
\kappa_{12\ell}
+
0.20A_i
+
0.25X_i
+
0.70
\log\left(
\frac{Y_{i8}^{\left(0\right)}+0.5}{3}
\right).
\end{align}
The quantities $\kappa_{8\ell}$ and $\kappa_{12\ell}$ are replicate-specific intercepts in the dropout models.

Within each replicate, $\kappa_{8\ell}$ was calibrated to give a mean Week~8 dropout probability of 0.35. After generating the Week~8 dropout indicators, we calibrated $\kappa_{12\ell}$ among the remaining participants so that those missing only at Week~12 comprised 0.20 of the original randomized sample in expectation. Equivalently, if $\overline{G}_{8\ell}$ is the realized Week~8 dropout proportion, the conditional target among those still at risk was $0.20/\left(1-\overline{G}_{8\ell}\right)$. Approximately 35\% of outcomes were therefore expected to be missing at Week~8 and 55\% at Week~12, with none missing at Week~4.

Let $\omega_{\mathrm{true}}$ denote the common value of the sensitivity parameter used from the trigger onward to generate outcomes for participants in the active group. We considered three post-trigger outcome mechanisms: $\omega_{\mathrm{true}} = 0$ under assigned arm continuation, $\omega_{\mathrm{true}} = 0.5$ under the intermediate rule, and $\omega_{\mathrm{true}} = 1$ under J2R. Outcomes in the reference arm were generated from the reference arm margin in every scenario. We then fitted the imputation model to the retained counts and completed the missing outcomes under each of the same three rules.

Let $g$ index the mechanism used to generate outcomes after the trigger. We included $R_{2g}=150$ replicates that met the convergence criteria for each mechanism, giving 450 replicates in total. These represented 150 of 151 attempts under assigned arm continuation, 150 of 152 under the intermediate rule, and all 150 attempts under J2R. The three excluded replicates did not meet the convergence criteria after the prespecified fitting attempts.

For every included replicate, we selected 50 approximately equally spaced draws from the combined post-warmup draws across the three chains and generated one completed dataset per draw under each imputation rule. Each dataset was analyzed with a working NB model. Let
\begin{equation}
\mu_{ij}^{\mathrm{ana}}
=
\E\left(
Y_{ij}
\middle|
A_i,X_i,e_{ij}
\right).
\end{equation}
The working mean model was
\begin{equation}
\log\left(
\mu_{ij}^{\mathrm{ana}}
\right)
=
\log\left(e_{ij}\right)
+
\gamma_{0j}
+
\gamma_j A_i
+
\gamma_X X_i,
\end{equation}
where $\gamma_{0j}$ is the intercept at visit $j$, $\gamma_j$ is the treatment log rate ratio at that visit, and $\gamma_X$ is the baseline covariate coefficient common across visits. The variance function was
\begin{equation}
\Var\left(
Y_{ij}
\middle|
A_i,X_i,e_{ij}
\right)
=
\mu_{ij}^{\mathrm{ana}}
+
\alpha_{\mathrm{ana}}
\left(
\mu_{ij}^{\mathrm{ana}}
\right)^2.
\end{equation}
The analysis used a common overdispersion parameter $\alpha_{\mathrm{ana}}$ across visits and treated repeated outcomes from the same participant as independent. We applied the same working model to the complete data before missingness was imposed and to every completed dataset.

Across the three generating mechanisms, the mean proportion missing from Week~8 onward ranged from 34.6\% to 35.4\%. The mean proportion missing only at Week~12 ranged from 20.0\% to 20.3\%, giving overall Week~12 missingness of 55.0\% to 55.3\%. \supptabref{supptab:simulation-achieved-missingness} gives results by mechanism and treatment group.

Let $r$ index the imputation rule. For each included replicate, we also fitted the analysis model to the complete outcome vector before imposing missingness. At visit $j$, the paired difference was
\begin{equation}
\Delta_{g\ell jr}
=
\widehat{\gamma}_{g\ell jr}^{\mathrm{MI}}
-
\widehat{\gamma}_{g\ell j}^{\mathrm{full}},
\end{equation}
where the superscript $\mathrm{MI}$ denotes multiple imputation, $\widehat{\gamma}_{g\ell jr}^{\mathrm{MI}}$ is the Rubin-pooled log rate ratio after imputation, and $\widehat{\gamma}_{g\ell j}^{\mathrm{full}}$ is the estimate from the corresponding complete data. The comparison is therefore with the treatment effect estimated from the complete data in the same replicate, not with the fixed generating coefficient $\tau_j$. We summarized agreement using the mean paired difference and paired RMSE,
\begin{equation}
\overline{\Delta}_{gjr}
=
\frac{1}{R_{2g}}
\sum_{\ell=1}^{R_{2g}}\Delta_{g\ell jr},
\qquad
\mathrm{RMSE}_{gjr}^{\mathrm{paired}}
=
\sqrt{\frac{1}{R_{2g}}
\sum_{\ell=1}^{R_{2g}}\Delta_{g\ell jr}^{2}}.
\end{equation}
We also report the empirical SD of the paired differences and the Monte Carlo standard error of $\overline{\Delta}_{gjr}$.

\autoref{tab:simulation-rbmi-performance} compares treatment effect estimates across the included replicates. The table focuses on Weeks~8 and 12, where outcomes were missing; Week~4 and additional paired summaries appear in \supptabref{supptab:rbmi-performance-full}. At both affected visits, the imputation rule matching the generating mechanism had the smallest absolute mean paired difference and, generally, the smallest paired RMSE. When the rule did not match, estimates shifted in the expected direction as the assumption moved from assigned arm continuation toward J2R. \supptabref{supptab:imputation-behavior} provides additional summaries of the imputed counts.

\begin{table}[!htbp]
\centering
\caption{Agreement between treatment effect estimates after reference-based imputation and the corresponding estimates from the complete data at visits affected by missingness.}
\label{tab:simulation-rbmi-performance}
\small
\setlength{\tabcolsep}{2pt}
\begin{tabular}{
@{}
>{\raggedright\arraybackslash}p{4cm}
>{\raggedright\arraybackslash}p{2.5cm}
cccccc
@{}
}
\toprule
\multirow[b]{2}{4cm}{\raggedright\textbf{Post-trigger outcome mechanism}}
&
\multirow[b]{2}{2.5cm}{\raggedright\textbf{Imputation rule}}
&
\multicolumn{3}{c}{\textbf{Week~8}}
&
\multicolumn{3}{c}{\textbf{Week~12}}
\\
\cmidrule(lr){3-5}
\cmidrule(lr){6-8}
&
&
\textbf{Mean diff.}
&
\textbf{MCSE}
&
\textbf{Paired RMSE}
&
\textbf{Mean diff.}
&
\textbf{MCSE}
&
\textbf{Paired RMSE}
\\
\midrule
Assigned arm cont. & Assigned arm & -0.012 & 0.012 & 0.147 & -0.019 & 0.018 & 0.225 \\
Assigned arm cont. & Intermediate & 0.092 & 0.010 & 0.152 & 0.152 & 0.014 & 0.226 \\
Assigned arm cont. & J2R & 0.212 & 0.010 & 0.247 & 0.344 & 0.013 & 0.378 \\
\addlinespace
Intermediate & Assigned arm & -0.073 & 0.013 & 0.172 & -0.136 & 0.018 & 0.262 \\
Intermediate & Intermediate & 0.020 & 0.011 & 0.134 & 0.013 & 0.014 & 0.168 \\
Intermediate & J2R & 0.129 & 0.012 & 0.192 & 0.180 & 0.013 & 0.240 \\
\addlinespace
J2R & Assigned arm & -0.189 & 0.014 & 0.257 & -0.319 & 0.022 & 0.419 \\
J2R & Intermediate & -0.093 & 0.012 & 0.173 & -0.160 & 0.016 & 0.257 \\
J2R & J2R & 0.018 & 0.011 & 0.139 & 0.018 & 0.014 & 0.170 \\
\bottomrule
\end{tabular}
\begin{flushleft}
\footnotesize
\justifying
Mean diff. is the average within-replicate difference between the pooled log rate ratio after imputation and the estimate from the corresponding complete data. MCSE is the Monte Carlo standard error of the mean difference. Paired RMSE is the root mean square of these differences. In the mechanism column, Assigned arm cont. denotes outcomes generated under assigned arm continuation. In the imputation rule column, Assigned arm denotes imputation under assigned arm continuation with $\omega_j=0$; Intermediate uses $\omega_j=0.5$; and J2R uses $\omega_j=1$.
\end{flushleft}
\end{table}

\section{Application to the motivating trial}
\label{sec:application}

We applied the method to the placebo and mirabegron 50~mg groups in the overactive bladder trial. The post-baseline outcome vector contained incontinence episode counts at Weeks~4, 8, and 12. Baseline was used as a covariate rather than included as an outcome in the copula. For an observed post-baseline count, we retained the recorded exposure; for a missing count, we set the exposure to 3 days, the intended diary window. Log exposure was an offset in both the imputation and analysis models.

The first missing scheduled post-baseline count served as the operational trigger. Because missingness was monotone, $\mis_i^{\pre}=\emptyset$ for every participant and completion began at the first missing visit.

\subsection{Imputation model}

The imputation model was fitted to the post-baseline count vector
\begin{equation}
\left(
Y_{i4}, Y_{i8}, Y_{i12}
\right)^\top,
\end{equation}
where the subscripts are study weeks. The marginal mean model had separate intercepts and treatment effects at each visit, covariate effects common across visits, and a log-exposure offset.

The baseline covariate was defined as a centered baseline log rate. Specifically,
\begin{equation}
L_i
=
\log\left(
\frac{Y_{i0} + 0.5}{e_{i0}}
\right),
\end{equation}
where $Y_{i0}$ is the baseline count, $e_{i0}$ is the number of valid days contributing to the baseline diary window, and adding 0.5 ensures that the baseline log rate is defined when $Y_{i0}=0$. The covariate included in the marginal mean model was
\begin{equation}
B_i
=
L_i
-
\bar{L},
\quad
\bar{L}
=
\frac{1}{n}
\sum_{i = 1}^{n}
L_i.
\end{equation}
The value $B_i = 0$ corresponds to the sample mean baseline log rate. The column vector $W_i$ contains indicators for age $\geq 65$ years, male sex, and Eastern Europe. The respective reference categories are age $<65$ years, female sex, and regions other than Eastern Europe.

The fitted marginal model can be written as
\begin{equation}
\log\left(\mu_{ij}^{a}\right)
=
\log\left(e_{ij}\right)
+
\eta_j
+
\tau_j a
+
\beta_{\mathrm{base}} B_i
+
W_i^\top\zeta,
\quad
j \in \left\{4, 8, 12\right\}.
\end{equation}
Here, $a = 0$ denotes placebo and $a = 1$ denotes mirabegron 50~mg. The coefficient $\beta_{\mathrm{base}}$ corresponds to the centered baseline log rate, and $\zeta$ contains the common effects of age group, sex, and region. The corresponding elements are denoted by $\zeta_{\mathrm{age}}$, $\zeta_{\mathrm{male}}$, and $\zeta_{\mathrm{EE}}$, where EE denotes Eastern Europe.

Under this parameterization, $\eta_j$ is the log daily rate for placebo at visit $j$ for a participant with the sample mean baseline log rate and reference values of the other covariates. Conditional on the same covariates, $\tau_j$ is the log rate ratio comparing mirabegron 50~mg with placebo at visit $j$.

This model is a parsimonious version of the general model in \autoref{sec:methods}, with covariate effects shared across post-baseline visits. Overdispersion parameters remained visit-specific. An AR(1) latent Gaussian correlation matrix described longitudinal dependence according to scheduled visit order, so $\rho$ is the latent correlation between adjacent visits.

Posterior computation used the randomized PIT augmentation, Metropolis-within-Gibbs sampler, and priors described in \autoref{subsec:estimation-implementation}. Four independent chains ran in parallel. Each chain had 5,000 warmup iterations followed by 5,000 retained draws without thinning, for 20,000 post-warmup draws in total. Proposal scales were adapted every 200 warmup iterations and then fixed. We assessed convergence and Monte Carlo accuracy using the PSRF \citep{gelman1992inference,brooks1998general}, approximate effective sample sizes, acceptance rates for parameter blocks, and aggregate acceptance rates for the PIT variables.

\supptabref{supptab:application-posterior-summary} reports selected posterior summaries from the imputation model: conditional daily rates for the reference covariate profile, rate ratios comparing mirabegron 50~mg with placebo, covariate effects, overdispersion by visit, and the AR(1) latent Gaussian correlation.

\subsection{Posterior predictive residual checks}
\label{subsec:application-ppc-residuals}

We used marginal posterior predictive PIT residuals to assess calibration of the fitted NB margins for recorded outcomes. At each retained posterior draw, we generated one complete post-baseline trajectory for each participant from the fitted NB Gaussian copula model, conditional on treatment, baseline covariates, and exposure at each visit. Replicated counts were compared with recorded counts using the randomized PIT method for discrete responses implemented in \texttt{DHARMa} \citep{HARTIG2021A}. The corresponding tests assessed residual uniformity, dispersion, excess zeros, and the number of observations outside the posterior predictive simulation envelope.

The model was fitted to the same outcomes used in this diagnostic, and the diagnostic included repeated observations from each participant. We therefore treat its $p$-values as nominal descriptive summaries rather than calibrated frequentist tests. In addition, the replicated trajectories were not conditioned on each participant's recorded post-baseline counts, and we retained the actual missingness pattern rather than simulating a dropout process. Consequently, the diagnostic assesses marginal posterior predictive calibration where outcomes were recorded. It does not directly validate the conditional copula dependence or account for selection into later observed visits that depends on outcomes. \appref{app:ppc-residual-check} summarizes the procedure.

\suppfigref{suppfig:application-ppc-residuals} compares empirical quantiles of the marginal posterior predictive PIT residuals with their expected uniform quantiles. The points followed the 45-degree line over much of the distribution, with modest departures in the middle and upper quantiles. The nominal one-sample Kolmogorov-Smirnov test indicated a departure from exact uniformity: the maximum absolute difference from the $\mathrm{Uniform}\left(0,1\right)$ distribution function was $D = 0.0634$ ($p < 0.001$). There was no evidence of underdispersion or overdispersion (dispersion statistic $= 0.965$, $p = 0.982$) or of excess zeros (observed-to-expected zero ratio $= 1.102$, $p = 0.188$). Three observations lay outside the posterior predictive simulation envelope, giving a nominal $p < 0.001$ in an exact binomial calculation with an approximate expected frequency. Overall, the diagnostics suggested modest nonuniformity and a few localized envelope departures, but no residual underdispersion, overdispersion, or excess zeros.

\subsection{Reference-based rules and completed datasets}

We considered three imputation rules at each post-baseline visit: assigned arm continuation with $\omega_j = 0$, the intermediate rule with $\omega_j = 0.5$, and J2R with $\omega_j = 1$. Under the intermediate rule, the log marginal mean lies halfway between the corresponding log marginal means under the assigned and reference arms.

Posterior summaries and predictive checks used all 20,000 post-warmup draws. For the treatment effect analysis, we selected 50 approximately equally spaced joint draws from the combined post-warmup draws across the four chains and generated one completed dataset per draw under each rule using \autoref{alg:conditional-rbmi}, reusing the sampled missing $U$ values across rules. Treatment effect estimates were combined using Rubin's rules as described in \autoref{subsec:rubin-pooling}.

\subsection{Analysis of observed and completed data}

We analyzed the observed data and each completed dataset with the same working NB generalized linear mixed model (GLMM), fitted in \texttt{glmmTMB} \citep{brooks2017glmmtmb}. The responses were incontinence episode counts at Weeks~4, 8, and 12. The model had a log link, an offset for log exposure, and fixed effects for visit, treatment by visit, centered baseline log rate, age group, sex, and region.

The conditional mean model can be written as
\begin{equation}
\log\left(\mu_{ij}^{\mathrm{ana}}\right)
=
\log\left(e_{ij}\right)
+
\delta_j
+
\gamma_j A_i
+
\beta_{\mathrm{base}}^{\mathrm{ana}} B_i
+
W_i^\top\zeta_{\mathrm{ana}}
+
b_i,
\quad
j \in \left\{4, 8, 12\right\},
\end{equation}
where $B_i$ is the centered baseline log rate and $W_i$ is the vector of baseline covariates defined above. At visit $j$, $\delta_j$ is the intercept for placebo and $\gamma_j$ is the log rate ratio comparing mirabegron 50~mg with placebo. The coefficients $\beta_{\mathrm{base}}^{\mathrm{ana}}$ and $\zeta_{\mathrm{ana}}$ correspond to $B_i$ and $W_i$, respectively, and $b_i$ is a subject-specific random intercept shared across visits.

The random intercepts were modeled as
\begin{equation}
b_i
\sim
\Normal\left(0, \sigma_b^2\right),
\end{equation}
independently across participants. This is specified as \texttt{(1 | id)} in \texttt{glmmTMB}. The random intercept induces dependence among repeated counts from the same participant. It provides a parsimonious working structure for the analysis and is not intended to reproduce the latent Gaussian dependence used for imputation. The model converged for the observed data and all completed datasets.

The analysis model used the quadratic NB variance function \citep{hardin2007generalized},
\begin{equation}
\Var\left(
Y_{ij}
\middle|
b_i, A_i, B_i, W_i, e_{ij}
\right)
=
\mu_{ij}^{\mathrm{ana}}
+
\alpha_{\mathrm{ana}}
\left(\mu_{ij}^{\mathrm{ana}}\right)^2.
\end{equation}
Within each fitted analysis model, $\alpha_{\mathrm{ana}}$ was common across visits. The imputation model, by contrast, had a separate overdispersion parameter at each visit.

At each visit, we summarized the conditional treatment effect by the fixed effect estimate $\gamma_j$ and rate ratio $\exp\left(\gamma_j\right)$. All $M=50$ completed datasets were analyzed successfully under every rule. We combined their treatment effect estimates using the Rubin procedure in \autoref{subsec:rubin-pooling}; confidence intervals from the observed data used a standard normal critical value.

\begin{figure}[ht]
\centering
\includegraphics[width=0.7\textwidth]{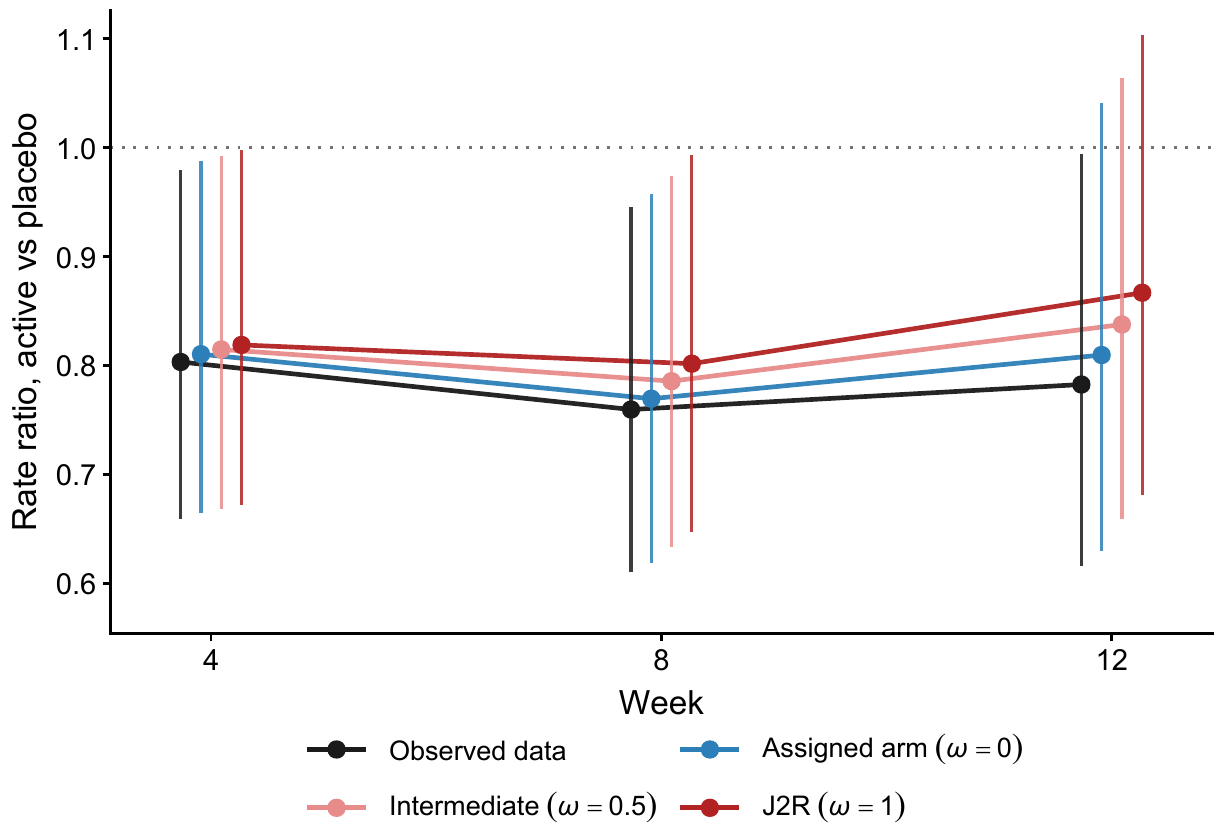}
\caption{
Rate ratios comparing mirabegron 50~mg with placebo by visit, based on the observed data and each reference-based analysis. Intervals are 95\% confidence intervals for the observed data and Rubin-pooled 95\% intervals after imputation. For missing outcomes in the mirabegron group from the trigger onward, $\omega_j = 0$ denotes assigned arm continuation, $\omega_j = 0.5$ the intermediate rule, and $\omega_j = 1$ J2R.
}
\label{fig:application-treatment-effects-by-visit}
\end{figure}

\autoref{fig:application-treatment-effects-by-visit} shows the results from the observed data and the three reference-based analyses. At each post-baseline visit, every analysis gave a rate ratio below 1, indicating that the model estimated a lower incontinence episode rate with mirabegron 50~mg than with placebo. The rules differed most at Week~12, where missingness was also greatest. Based on the observed data, the Week~12 rate ratio was 0.782 with a 95\% confidence interval of $\left(0.616, 0.994\right)$. The Rubin-pooled estimates were 0.810 with a 95\% interval of $\left(0.630, 1.041\right)$ under assigned arm continuation, 0.838 with $\left(0.659, 1.064\right)$ under the intermediate rule, and 0.867 with $\left(0.681, 1.103\right)$ under J2R. Moving from assigned arm continuation to J2R attenuated the estimated effect toward the null without changing its direction. All three Week~12 intervals after imputation included 1.

\section{Discussion}
\label{sec:discussion}

We developed a reference-based sensitivity analysis for repeated counts recorded at scheduled visits. NB margins preserve integer-valued imputations, accommodate overdispersion and exposure offsets, and allow the retained treatment effect to be specified on the log rate scale. A Gaussian copula supplies the working dependence model needed to condition imputations on retained counts. Randomized PIT augmentation provides latent variables compatible with the discrete margins during posterior computation.

The copy-reference implementation in \texttt{gsDesignNB} uses an NB GLMM to combine fixed-effect predictions from the reference group with a multiplier derived from subject-specific random effects. Our method instead applies the sensitivity rule to the marginal mean and uses a Gaussian copula for dependence and conditional imputation. Even when the two approaches imply the same post-trigger mean, they can produce different conditional distributions for the missing counts because subject-specific random effects and a Gaussian copula induce within-participant dependence differently.

The two simulation studies had different aims. With fully observed outcomes, the fitted model recovered the generating marginal and dependence parameters with little bias, and posterior SDs generally agreed with the empirical SDs of the posterior means. Coverage was 0.94 to 0.96 for most parameters and 0.92 for $\tau_4$. The simulation with incomplete data compared treatment effect estimates after imputation with estimates from the corresponding complete data before missingness was imposed. Recovery at affected visits was best when the imputation rule matched the mechanism used to generate outcomes after the trigger. Mismatched rules shifted the estimates in the direction implied by their assumptions. The second study was designed to compare paired point estimates within replicates under specific generating settings. It did not evaluate the operating characteristics of Rubin-pooled intervals.

In the overactive bladder application, the marginal posterior predictive residuals showed modest nonuniformity, with three points outside the simulation envelope, but no evidence of underdispersion, overdispersion, or excess zeros. The separate diagnostic fitted to the observed counts also found no evidence of excess zeros beyond those accommodated by an NB model. Across the sensitivity analyses, the estimated treatment effect retained the same direction. At Week~12, the treatment contrast moved toward the null, as expected, when the assumption changed from assigned arm continuation to J2R.

As with other reference-based analyses, the distribution of missing outcomes after the trigger cannot be identified from the observed data alone. The prespecified values of $\omega$ should therefore be interpreted as sensitivity assumptions, rather than as treatment retention parameters estimated from the data. The NB margins and Gaussian copula determine how each assumption is translated into a conditional imputation distribution. Consequently, results remain dependent on the adequacy of both the marginal and dependence models.

The reference-based rules developed here do not enforce information anchoring, under which the proportion of information lost because of missing data is approximately preserved between the primary and sensitivity analyses \citep{cro2019information,atkinson2021information}. Unlike multivariate normal reference-based imputation with a common covariance structure, an NB margin has a variance that depends on its mean. Changing the marginal mean under a rule can therefore change the predictive variance of the count, even if overdispersion is fixed. Similarly, keeping the latent Gaussian correlation fixed does not keep the covariance of the observed counts fixed across rules. Rubin-pooled intervals should consequently be interpreted as conventional multiple imputation intervals under the specified imputation and analysis procedures, without a formal guarantee of information anchoring. Formal evaluation of information anchoring for this method of imputing counts remains a topic for future work.

Possible extensions include other discrete margins and more flexible copula structures. If excess zeros remain after accounting for NB overdispersion, the same copula and randomized PIT approach could use zero-inflated NB or hurdle margins. Each reference-based rule would then need to specify how both the count component and the probability of a zero change.

In summary, the method extends RBMI to repeated count endpoints while retaining their natural count scale and accommodating unequal exposure. It is intended for counts or rates recorded at scheduled visits, rather than recurrent event processes observed continuously over follow-up.

\section*{Author contributions}

Divan A. Burger conceptualized the study, developed the methodology, implemented the computational approach, performed the analyses, interpreted the results, and drafted the manuscript. Emmanuel Lesaffre and Reynaldo Martina contributed to the methodological development, interpretation of the results, and critical revision of the manuscript. All authors reviewed and approved the final manuscript.

\section*{Funding}

This work is based on research supported by the National Research Foundation of South Africa (Grant number 132383). Opinions expressed and conclusions arrived at are those of the authors and are not necessarily to be attributed to the National Research Foundation.

\section*{Conflicts of interest}

Divan A. Burger is an employee of Cytel Inc. The other authors declare no conflicts of interest.

\section*{Ethics statement}

The application was a secondary analysis of data from a completed randomized clinical trial. The dataset contained no directly identifying participant information. This methodological analysis involved no additional participant contact, intervention, or data collection. The original mirabegron trial, conducted in Europe and Australia (ClinicalTrials.gov identifier NCT00689104), was approved by the relevant local institutional review boards or independent ethics committees and was conducted in accordance with Good Clinical Practice and the Declaration of Helsinki. All patients provided written informed consent \citep{khullar2013phase3}.

\section*{Disclaimer}
    
The views expressed in this article are those of the authors and do not necessarily reflect the views or policies of Cytel Inc.

\section*{Data and code availability statement}

The clinical trial data used in the application are confidential and cannot be shared publicly. They were used for this methodological illustration under the applicable data use restrictions. Code for the simulations and application analysis is available at \url{https://github.com/DABURGER1/RBMI-NB-Counts}; the repository contains no individual participant data.

\ifArxivVersion
  % arXiv version: read the complete bibliography saved from Overleaf.
  % Keep bib.bbl in the same folder as this main TeX file.

\else
  % Original version: generate the bibliography from bib.bib.
  \bibliographystyle{unsrtnat}
  \bibliography{bib}
\fi

\appendix

\setcounter{algorithm}{0}
\renewcommand{\thealgorithm}{\Alph{algorithm}}
\providecommand{\theHalgorithm}{\thealgorithm}
\renewcommand{\theHalgorithm}{app.\Alph{algorithm}}
\renewcommand{\algorithmautorefname}{Appendix Algorithm}

\section*{Appendix}

\section{Randomized probability integral transform augmentation}
\label{app:pit-augmentation}

For a fixed vector $\theta$ of marginal and copula parameters on their natural scales, write $R = R\left(\psi\right)$. Let $c_{R_{\obs_i\obs_i}}$ be the Gaussian copula density associated with $R_{\obs_i\obs_i}$. On the uniform scale, the probability of the retained counts is
\begin{equation}
\Pr\left(Y_{i,\obs_i} = y_{i,\obs_i} \middle| A_i, X_i, e_i, \theta\right)
=
\int_{\prod_{j \in \obs_i}
\left(
F_{ij}\left(y_{ij} - 1\right),
F_{ij}\left(y_{ij}\right)
\right]}
c_{R_{\obs_i\obs_i}}\left(q_{i,\obs_i}\right)
d q_{i,\obs_i}.
\end{equation}
For $j \in \obs_i$, set
\begin{equation}
q_{ij}
=
F_{ij}\left(y_{ij} - 1\right)
+
v_{ij}
\pi_{ij}\left(y_{ij}\right),
\quad
0 < v_{ij} < 1.
\end{equation}
Then
\begin{equation}
d q_{i,\obs_i}
=
\prod_{j \in \obs_i}
\pi_{ij}\left(y_{ij}\right)
d v_{i,\obs_i}.
\end{equation}
Therefore,
\begin{equation}
\Pr\left(Y_{i,\obs_i} = y_{i,\obs_i} \middle| A_i, X_i, e_i, \theta\right)
=
\int_{\left(0,1\right)^{\left|\obs_i\right|}}
c_{R_{\obs_i\obs_i}}\left(q_{i,\obs_i}\right)
\prod_{j \in \obs_i}
\pi_{ij}\left(y_{ij}\right)
d v_{i,\obs_i}.
\end{equation}
Let $v_{i,\obs_i}=\left(v_{ij}:j\in\obs_i\right)^\top$ be a realization of the auxiliary variables $V_{i,\obs_i}=\left(V_{ij}:j\in\obs_i\right)^\top$. The vector $\theta$ contains the marginal and copula parameters on their natural scales and determines $F_{ij}$, $\pi_{ij}$, and $R$. Conditional on a fixed $\theta$ and the retained count vector, the density of $V_{i,\obs_i}$ is proportional to the augmented likelihood:
\begin{equation}
p\left(v_{i,\obs_i} \middle| y_{i,\obs_i}, A_i, X_i, e_i, \theta\right)
=
\frac{
L_i^{\aug}\left(\theta,v_{i,\obs_i}\right)
}{
\int_{\left(0, 1\right)^{\left|\obs_i\right|}}
L_i^{\aug}\left(\theta,\tilde{v}_{i,\obs_i}\right)
\prod_{j \in \obs_i}
d \tilde{v}_{ij}
},
\end{equation}
where $\tilde{v}_{i,\obs_i}$ is a dummy integration variable over $\left(0,1\right)^{\left|\obs_i\right|}$. The product uniform measure is the base measure for the augmentation, but the augmented likelihood determines the conditional distribution of the PIT variables.

Integrating out the auxiliary PIT variables recovers the exact rectangle probability for the retained counts:
\begin{equation}
\Pr\left(Y_{i,\obs_i} = y_{i,\obs_i} \middle| A_i, X_i, e_i, \theta\right)
=
\int_{\left(0, 1\right)^{\left|\obs_i\right|}}
L_i^{\aug}\left(\theta,v_{i,\obs_i}\right)
\prod_{j \in \obs_i}
d v_{ij}.
\end{equation}

\section{Conditional distribution under an imputation rule}
\label{app:conditional-distribution}

For a fixed vector $\theta$ of marginal and copula parameters on their natural scales, write $R = R\left(\psi\right)$. We treat the realized trigger index $D_i$ as fixed and omit it from the conditioning arguments below. It determines the partition of $\mis_i$ into $\mis_i^{\pre}$ and $\mis_i^{\post}$. For $j \in \obs_i$, the latent interval for a retained count is
\begin{equation}
\mathcal{B}_{ij}^{\obsval}\left(y\right)
=
\left(
\Phi^{-1}\left[
F_{ij}^{A_i}\left(y - 1\right)
\right],
\Phi^{-1}\left[
F_{ij}^{A_i}\left(y\right)
\right]
\right].
\end{equation}
For $j \in \mis_i$, the latent interval under rule $r$ is
\begin{equation}
\mathcal{B}_{ij}^{r}\left(y\right)
=
\left(
\Phi^{-1}\left[
F_{ij}^{r}\left(y - 1\right)
\right],
\Phi^{-1}\left[
F_{ij}^{r}\left(y\right)
\right]
\right].
\end{equation}
Let
\begin{equation}
\mathcal{B}_{i,\obs_i}^{\obsval}
=
\prod_{j \in \obs_i}
\mathcal{B}_{ij}^{\obsval}\left(y_{ij}\right)
\end{equation}
denote the rectangle for the retained counts at the conditioning visits.

The joint probability under rule $r$ uses margins for the assigned arm at visits in $\obs_i$ and margins determined by the rule at visits in $\mis_i$:
\begin{equation}
p_r\left(
y_{i,\obs_i},
y_{i,\mis_i}
\middle|
A_i,
X_i,
e_i
\right)
=
\int_{
\prod_{j \in \obs_i}
\mathcal{B}_{ij}^{\obsval}\left(y_{ij}\right)
\times
\prod_{j \in \mis_i}
\mathcal{B}_{ij}^{r}\left(y_{ij}\right)
}
\phi_J\left(z_i;0,R\right)
d z_i.
\end{equation}
Here, $\phi_J\left(\cdot;0,R\right)$ denotes the $J$-variate normal density with mean zero and correlation matrix $R$.

Under rule $r$, the conditional distribution of counts in $\mis_i$ given those in $\obs_i$ is
\begin{equation}
p_r\left(y_{i,\mis_i} \middle| y_{i,\obs_i}, A_i, X_i, e_i\right)
=
\frac{
p_r\left(y_{i,\obs_i}, y_{i,\mis_i} \middle| A_i, X_i, e_i\right)
}{
p\left(y_{i,\obs_i} \middle| A_i, X_i, e_i\right)
}.
\end{equation}
The denominator is the marginal probability of the retained counts. It can be obtained by summing the joint probability under rule $r$ over all possible count vectors in $\mis_i$:
\begin{equation}
p\left(y_{i,\obs_i} \middle| A_i, X_i, e_i\right)
=
\sum_{\tilde{y}_{i,\mis_i} \in \Nzero^{\left|\mis_i\right|}}
p_r\left(y_{i,\obs_i}, \tilde{y}_{i,\mis_i} \middle| A_i, X_i, e_i\right),
\end{equation}
where $\tilde{y}_{i,\mis_i}$ ranges over every nonnegative integer count vector for visits in $\mis_i$. Although each summand is defined under rule $r$, the resulting marginal probability does not depend on the rule. For any fixed rule, the latent regions for all possible count vectors in $\mis_i$ partition the full latent space at those visits. Summing over them integrates out the missing latent components and leaves the common marginal distribution of the retained counts:
\begin{equation}
\sum_{\tilde{y}_{i,\mis_i} \in \Nzero^{\left|\mis_i\right|}}
p_r\left(y_{i,\obs_i}, \tilde{y}_{i,\mis_i} \middle| A_i, X_i, e_i\right)
=
\int_{\mathcal{B}_{i,\obs_i}^{\obsval} \times \R^{\left|\mis_i\right|}}
\phi_J\left(z_i;0,R\right)
d z_i
=
\int_{\mathcal{B}_{i,\obs_i}^{\obsval}}
\phi_{\left|\obs_i\right|}
\left(
z_{i,\obs_i};0,R_{\obs_i\obs_i}
\right)
d z_{i,\obs_i}.
\end{equation}
Hence, the marginal distribution of counts in $\obs_i$ is the same under every reference-based rule, provided that the rule changes only the margins at visits in $\mis_i$ and leaves the margins at visits in $\obs_i$ and the fitted latent Gaussian correlation matrix unchanged.

\section{Pooling estimates from completed datasets}
\label{subsec:rubin-pooling}

Pooling is performed separately for each imputation rule and each scalar analysis estimand. Let $s_1,\ldots,s_M$ denote the selected posterior draws used to generate the $M$ completed datasets that were analyzed successfully. For rule $r$, let $\widehat{Q}_m^{\left(r\right)}$ be the estimate from the completed dataset generated using draw $s_m$, and let $U_m^{\left(r\right)}$ be its estimated variance. In this subsection, the symbols $\overline{Q}^{\left(r\right)}$, $\overline{U}^{\left(r\right)}$, $B^{\left(r\right)}$, and $T^{\left(r\right)}$ are used in their conventional Rubin pooling sense. The pooled estimate, within-imputation variance, between-imputation variance, and total variance are
\begin{equation}
\begin{aligned}
\overline{Q}^{\left(r\right)}
&=
\frac{1}{M}
\sum_{m=1}^{M}
\widehat{Q}_m^{\left(r\right)},
\\
\overline{U}^{\left(r\right)}
&=
\frac{1}{M}
\sum_{m=1}^{M}
U_m^{\left(r\right)},
\\
B^{\left(r\right)}
&=
\frac{1}{M-1}
\sum_{m=1}^{M}
\left(
\widehat{Q}_m^{\left(r\right)}
-
\overline{Q}^{\left(r\right)}
\right)^2,
\\
T^{\left(r\right)}
&=
\overline{U}^{\left(r\right)}
+
\left(
1+\frac{1}{M}
\right)
B^{\left(r\right)}.
\end{aligned}
\end{equation}
When $B^{\left(r\right)}>0$, the usual Rubin approximation gives the finite-$M$ degrees of freedom:
\begin{equation}
\nu^{\left(r\right)}
=
\left(M-1\right)
\left[
1+
\frac{
\overline{U}^{\left(r\right)}
}{
\left(1+M^{-1}\right)
B^{\left(r\right)}
}
\right]^2.
\end{equation}
A two-sided 95\% interval is
\begin{equation}
\overline{Q}^{\left(r\right)}
\pm
t_{\nu^{\left(r\right)},\,0.975}
\sqrt{
T^{\left(r\right)}
},
\end{equation}
where $t_{\nu^{\left(r\right)},\,0.975}$ denotes the 97.5th percentile of a Student's $t$ distribution with $\nu^{\left(r\right)}$ degrees of freedom.

When $B^{\left(r\right)}=0$, we treat $\nu^{\left(r\right)}$ as infinite and use the standard normal critical value. We do not apply the small-sample degrees-of-freedom correction proposed by Barnard and Rubin \citep{barnard1999miscellanea}, which incorporates the complete data degrees of freedom. In the treatment analyses, $\widehat{Q}_m^{\left(r\right)}$ is the estimated log rate ratio and $U_m^{\left(r\right)}$ is the square of its model-based standard error. Pooling is performed on the log rate ratio scale, and the pooled estimate and interval limits are then exponentiated for presentation as rate ratios.

\section{Posterior sampler details}
\label{app:posterior-sampler}

This appendix gives further details of the Metropolis-within-Gibbs sampler in \autoref{subsec:estimation-implementation}. The cycle in \appalgref{appalg:posterior-sampler} alternates among the unconstrained marginal parameters, the unconstrained copula parameter, and PIT variables for each participant. The proposal covariance $S_b^{\mathrm{prop}}$ and PIT proposal variance $\sigma_V^2$ are tuning quantities. Posterior draws are transformed to their natural scales before use in the conditional imputation procedure in \mainalgref{alg:conditional-rbmi}.

\begin{algorithm}[ht]
\caption{Metropolis-within-Gibbs sampler for the posterior with randomized PIT augmentation}
\label{appalg:posterior-sampler}
\centering
\fbox{
\begin{minipage}{0.97\textwidth}
\footnotesize
\begin{enumerate}[leftmargin=1.75em, itemsep=2pt]
\item For each chain, initialize the computational parameter vector $\vartheta^{\left(0\right)}$ and the PIT variables
\begin{equation}
V_{\obs}^{\left(0\right)}
=
\left\{
V_{ij}^{\left(0\right)}:
j \in \obs_i,\ i = 1,\ldots,n
\right\}.
\end{equation}

\item Let $K_{\mathrm{MCMC}}$ denote the total number of Markov chain Monte Carlo (MCMC) iterations in that chain. At iteration $k=1,\ldots,K_{\mathrm{MCMC}}$, update the marginal model parameters in prespecified blocks. For block $b$, propose
\begin{equation}
\vartheta_b^{\dagger}
=
\vartheta_b^{\mathrm{cur}}
+
\varepsilon_b,
\quad
\varepsilon_b
\sim
\Normal\left(0,S_b^{\mathrm{prop}}\right),
\end{equation}
and accept or reject the proposal using the augmented log posterior
\begin{equation}
\ell_{\aug}\left(\vartheta,V_{\obs}\right).
\end{equation}

\item Update the unconstrained copula dependence coordinate $\psi_{\mathrm{raw}}$. For the AR(1) model used in the simulations and application, this means updating $\rho_{\mathrm{raw}}$ and transforming it to $\rho=2\Phi\left(\rho_{\mathrm{raw}}\right)-1$ before evaluating the copula contribution.

\item For each participant $i$ with $\obs_i \neq \emptyset$, update the PIT variables $V_{i,\obs_i}$ by proposing on the logit scale:
\begin{equation}
H_{i,\obs_i}^{\dagger}
=
H_{i,\obs_i}^{\mathrm{cur}}
+
\varepsilon_i,
\quad
\varepsilon_i
\sim
\Normal
\left(
0,
\sigma_V^2 I_{\left|\obs_i\right|}
\right),
\end{equation}
where
\begin{equation}
H_{ij}
=
\log
\left(
\frac{V_{ij}}{1 - V_{ij}}
\right),
\quad
j \in \obs_i.
\end{equation}
Here, $H_{i,\obs_i}^{\mathrm{cur}}$ is the current PIT vector on the logit scale for participant $i$. Transform the proposal back to $\left(0,1\right)$ and accept or reject it using the copula contribution for that participant and the log-Jacobian term.

\item During warmup, adapt the proposal scales for the parameter blocks and PIT variable updates using acceptance rates over short windows. Fix all proposal scales after warmup.

\item After warmup, retain every post-warmup joint draw from every chain and index the combined set as
\begin{equation}
\left(
\vartheta^{\left(s\right)},
V_{\obs}^{\left(s\right)}
\right),
\quad
s = 1,\ldots,S.
\end{equation}

\item Transform each retained parameter draw to
\begin{equation}
\theta^{\left(s\right)}
=
h\left(\vartheta^{\left(s\right)}\right).
\end{equation}
Use the selected joint draws
\begin{equation}
\left(
\theta^{\left(s\right)},
V_{\obs}^{\left(s\right)}
\right)
\end{equation}
in \mainalgref{alg:conditional-rbmi} to generate completed count vectors under each reference-based rule.
\end{enumerate}
\end{minipage}
}
\end{algorithm}

\section{Posterior predictive residual check}
\label{app:ppc-residual-check}

\appalgref{appalg:ppc-residual-check} summarizes the posterior predictive residual check used in the application. The diagnostic is separate from the reference-based imputation procedure and assesses marginal calibration of the fitted NB margins where outcomes were recorded. It does not directly assess conditional dependence under the copula.
%\clearpage
\begin{algorithm}[ht]
\caption{Marginal posterior predictive PIT residual check for the fitted imputation model}
\label{appalg:ppc-residual-check}
\centering
\fbox{
\begin{minipage}{0.97\textwidth}
\footnotesize
\begin{enumerate}[leftmargin=1.75em, itemsep=2pt]
\item For each retained posterior draw $s=1,\ldots,S$, transform the computational vector to the marginal and copula parameters on their natural scales, $\theta^{\left(s\right)}=h\left(\vartheta^{\left(s\right)}\right)$.

\item For each participant $i$, condition on the observed treatment assignment $A_i$, baseline covariates $X_i$, and visit-specific exposure values $e_{ij}$.

\item Generate a new latent Gaussian trajectory
\begin{equation}
Z_i^{\mathrm{rep},\left(s\right)}
\sim
\Normal\left(0, R\left(\psi^{\left(s\right)}\right)\right).
\end{equation}

\item Transform the latent Gaussian values to uniforms,
\begin{equation}
U_{ij}^{\mathrm{rep},\left(s\right)}
=
\Phi\left(Z_{ij}^{\mathrm{rep},\left(s\right)}\right),
\quad
j=1,\ldots,J.
\end{equation}

\item Map each uniform through the fitted NB margin under the assigned arm,
\begin{equation}
Y_{ij}^{\mathrm{rep},\left(s\right)}
=
\inf
\left\{
y \in \Nzero:
F_{ij}^{A_i}\left(y;\theta^{\left(s\right)}\right)
\geq
U_{ij}^{\mathrm{rep},\left(s\right)}
\right\}.
\end{equation}

\item For $j \in \obs_i$, compare the replicated count $Y_{ij}^{\mathrm{rep},\left(s\right)}$ with the recorded count $Y_{ij}$ using randomized posterior predictive PIT residuals on the uniform scale.
\end{enumerate}
\end{minipage}
}
\end{algorithm}

\StartSupplement

\section{Simulation study details}
\label{supp:simulation-details}

This section provides further details and results for the simulations in \mainsecref{sec:simulation}.

\begin{table}[ht]
\centering
\caption{Data-generating parameter values used in the simulation studies.}
\label{supptab:simulation-dgp}
\small
\begin{tabular}{@{}lccc@{}}
\toprule
\textbf{Parameter} & \textbf{Week~4} & \textbf{Week~8} & \textbf{Week~12} \\
\midrule
Reference arm rate $\exp\left(\eta_j\right)$ & 1.715 & 1.540 & 1.448 \\
Treatment rate ratio $\exp\left(\tau_j\right)$, model recovery & 0.909 & 0.851 & 0.859 \\
Treatment rate ratio $\exp\left(\tau_j\right)$, simulation with incomplete data & 0.909 & 0.650 & 0.600 \\
Overdispersion $\alpha_j$ & 1.206 & 1.037 & 0.981 \\
\bottomrule
\end{tabular}
\begin{flushleft}
\footnotesize
\justifying
Reference arm rates are conditional on $X_i = 0$. For treatment arm $a$, the generating mean at visit $j$ was $\mu_{ij}^{a} = e_{ij}\exp\left(\eta_j + \tau_j a + \beta_X X_i\right)$, where $X_i \sim \Normal\left(0,1\right)$, $\beta_X = 0.600$, and $e_{ij} = 3$ for every participant and visit. Both simulations used the same overdispersion parameters $\alpha_j$ and AR(1) latent Gaussian correlation $\rho = 0.462$. The first study used the treatment rate ratios shown in the model recovery row. The simulation with incomplete data used the stronger effects shown in the following row at Weeks~8 and 12.
\end{flushleft}
\end{table}

Both simulations used $n=150$ participants randomized equally to the reference and active arms. In the simulation with incomplete data, we generated a latent Gaussian trajectory and copula uniforms for each participant. Counts under assigned arm continuation, with $\omega_{\mathrm{true}}=0$, were first generated at every visit and provided the preceding counts used in the dropout mechanism. After determining the trigger, we generated the final complete trajectory by setting $\omega_{\mathrm{true}}$ to $0$, $0.5$, or $1$ for the active arm from the trigger onward under assigned arm continuation, the intermediate rule, or J2R, respectively. In each scenario, the assigned arm trajectory used for the dropout mechanism and the final complete trajectory used the same copula uniforms, preserving each participant's latent longitudinal trajectory. We then set outcomes from the trigger onward to missing.

Each simulated dataset was fitted with three independent chains. The first attempt used 2,000 warmup iterations followed by 2,000 retained draws per chain, without thinning. If the monitored core parameters did not meet the convergence criteria of a maximum PSRF of 1.05 and a minimum approximate effective sample size of 100, we refitted the model with 4,000 warmup and 4,000 retained draws per chain, and then, if needed, 6,000 warmup and 6,000 retained draws per chain. Replicates that still failed these criteria after the third attempt were excluded. All posterior parameter summaries for each included replicate used every post-warmup draw from the final fitting attempt.

\newpage

\begin{table}[!htbp]
\centering
\caption{Recovery of individual model parameters with fully observed outcomes.}
\label{supptab:model-recovery-full}
\small
\begin{tabular}{@{}lccccccc@{}}
\toprule
\textbf{Parameter} & \textbf{True} & \textbf{Mean} & \textbf{Bias} & \textbf{RMSE} & \textbf{Emp. SD} & \textbf{Mean post. SD} & \textbf{Coverage} \\
\midrule
$\eta_{4}$ & 0.539 & 0.536 & -0.004 & 0.138 & 0.139 & 0.138 & 0.96 \\
$\eta_{8}$ & 0.432 & 0.435 & 0.003 & 0.132 & 0.132 & 0.129 & 0.96 \\
$\eta_{12}$ & 0.370 & 0.378 & 0.007 & 0.130 & 0.130 & 0.128 & 0.95 \\
$\tau_{4}$ & -0.095 & -0.113 & -0.018 & 0.204 & 0.203 & 0.194 & 0.92 \\
$\tau_{8}$ & -0.161 & -0.182 & -0.021 & 0.191 & 0.190 & 0.181 & 0.95 \\
$\tau_{12}$ & -0.152 & -0.174 & -0.022 & 0.184 & 0.183 & 0.181 & 0.95 \\
$\beta_X$ & 0.600 & 0.608 & 0.008 & 0.069 & 0.069 & 0.070 & 0.95 \\
$\alpha_{4}$ & 1.206 & 1.233 & 0.027 & 0.187 & 0.185 & 0.179 & 0.94 \\
$\alpha_{8}$ & 1.037 & 1.050 & 0.013 & 0.163 & 0.162 & 0.161 & 0.94 \\
$\alpha_{12}$ & 0.981 & 1.003 & 0.022 & 0.154 & 0.153 & 0.156 & 0.95 \\
$\rho$ & 0.462 & 0.454 & -0.008 & 0.051 & 0.051 & 0.051 & 0.94 \\
\bottomrule
\end{tabular}
\begin{flushleft}
\footnotesize
\justifying
Mean is the average posterior mean across included simulated datasets. Bias is Mean minus True. RMSE is the root mean square error of the posterior mean relative to the true value. Emp. SD is the empirical standard deviation of posterior means across simulated datasets. Mean post. SD is the mean posterior standard deviation. Coverage is the empirical coverage of the 95\% posterior credible interval.
\end{flushleft}
\end{table}

\begin{landscape}

\begin{table}[ht]
\centering
\caption{Achieved missingness in the simulation with incomplete data.}
\label{supptab:simulation-achieved-missingness}
\scriptsize
\setlength{\tabcolsep}{3pt}
\begin{tabular}{@{}llcccc@{}}
\toprule
\textbf{Post-trigger outcome mechanism} & \textbf{Group} & \textbf{Replicates} & \textbf{Missing from Week~8 onward} & \textbf{Missing at Week~12 only} & \textbf{Missing at Week~12 overall} \\
\midrule
Assigned arm continuation & Overall & 150 & 34.6 (3.4) & 20.3 (2.9) & 55.0 (4.5) \\
Assigned arm continuation & Reference & 150 & 33.0 (5.0) & 21.2 (4.2) & 54.2 (6.2) \\
Assigned arm continuation & Active & 150 & 36.2 (5.5) & 19.5 (4.1) & 55.7 (6.0) \\
\addlinespace
Intermediate & Overall & 150 & 34.9 (3.3) & 20.2 (2.7) & 55.1 (4.2) \\
Intermediate & Reference & 150 & 33.5 (5.2) & 20.8 (4.3) & 54.3 (6.5) \\
Intermediate & Active & 150 & 36.2 (4.9) & 19.6 (4.2) & 55.8 (5.6) \\
\addlinespace
J2R & Overall & 150 & 35.4 (3.4) & 20.0 (3.0) & 55.3 (4.2) \\
J2R & Reference & 150 & 34.2 (5.2) & 20.9 (4.6) & 55.1 (5.8) \\
J2R & Active & 150 & 36.5 (5.0) & 19.0 (4.5) & 55.5 (5.5) \\
\bottomrule
\end{tabular}
\begin{flushleft}
\footnotesize
\justifying
Entries are means (empirical standard deviations) of the percentages across included replicates. Replicates is the number of included replicates. Missing from Week~8 onward denotes missing counts at both Weeks~8 and 12, while Missing at Week~12 only denotes a recorded Week~8 count and a missing Week~12 count. Overall Week~12 missingness is the sum of these two categories. The nominal targets were 35\% missing from Week~8 onward and a further 20\% missing only at Week~12, giving 55\% overall missingness at Week~12.
\end{flushleft}
\end{table}

\newpage

\begin{table}[ht]
\centering
\caption{Agreement between treatment effect estimates after imputation and estimates from the complete data, by post-trigger outcome mechanism, imputation rule, and visit.}
\label{supptab:rbmi-performance-full}
\scriptsize
\begin{tabular}{@{}llccccccc@{}}
\toprule
\textbf{Post-trigger outcome mechanism} & \textbf{Imputation rule} & \textbf{Week} & \textbf{Mean complete} & \textbf{Mean MI} & \textbf{Mean diff.} & \textbf{Paired SD} & \textbf{Paired RMSE} & \textbf{MCSE} \\
\midrule
Assigned arm continuation & Assigned arm &  4 & -0.119 & -0.121 & -0.001 & 0.009 & 0.009 & 0.001 \\
Assigned arm continuation & Assigned arm &  8 & -0.428 & -0.440 & -0.012 & 0.147 & 0.147 & 0.012 \\
Assigned arm continuation & Assigned arm & 12 & -0.502 & -0.522 & -0.019 & 0.225 & 0.225 & 0.018 \\
Assigned arm continuation & Intermediate &  4 & -0.119 & -0.121 & -0.002 & 0.010 & 0.010 & 0.001 \\
Assigned arm continuation & Intermediate &  8 & -0.428 & -0.337 & 0.092 & 0.121 & 0.152 & 0.010 \\
Assigned arm continuation & Intermediate & 12 & -0.502 & -0.351 & 0.152 & 0.168 & 0.226 & 0.014 \\
Assigned arm continuation & J2R &  4 & -0.119 & -0.122 & -0.003 & 0.011 & 0.011 & 0.001 \\
Assigned arm continuation & J2R &  8 & -0.428 & -0.216 & 0.212 & 0.127 & 0.247 & 0.010 \\
Assigned arm continuation & J2R & 12 & -0.502 & -0.159 & 0.344 & 0.159 & 0.378 & 0.013 \\
\addlinespace
Intermediate & Assigned arm &  4 & -0.105 & -0.106 & -0.001 & 0.010 & 0.010 & 0.001 \\
Intermediate & Assigned arm &  8 & -0.331 & -0.404 & -0.073 & 0.157 & 0.172 & 0.013 \\
Intermediate & Assigned arm & 12 & -0.324 & -0.460 & -0.136 & 0.225 & 0.262 & 0.018 \\
Intermediate & Intermediate &  4 & -0.105 & -0.106 & -0.001 & 0.010 & 0.010 & 0.001 \\
Intermediate & Intermediate &  8 & -0.331 & -0.311 & 0.020 & 0.133 & 0.134 & 0.011 \\
Intermediate & Intermediate & 12 & -0.324 & -0.310 & 0.013 & 0.168 & 0.168 & 0.014 \\
Intermediate & J2R &  4 & -0.105 & -0.106 & -0.001 & 0.011 & 0.011 & 0.001 \\
Intermediate & J2R &  8 & -0.331 & -0.203 & 0.129 & 0.143 & 0.192 & 0.012 \\
Intermediate & J2R & 12 & -0.324 & -0.144 & 0.180 & 0.159 & 0.240 & 0.013 \\
\addlinespace
J2R & Assigned arm &  4 & -0.096 & -0.097 & -0.001 & 0.010 & 0.010 & 0.001 \\
J2R & Assigned arm &  8 & -0.225 & -0.414 & -0.189 & 0.174 & 0.257 & 0.014 \\
J2R & Assigned arm & 12 & -0.171 & -0.490 & -0.319 & 0.272 & 0.419 & 0.022 \\
J2R & Intermediate &  4 & -0.096 & -0.097 & -0.001 & 0.010 & 0.010 & 0.001 \\
J2R & Intermediate &  8 & -0.225 & -0.318 & -0.093 & 0.146 & 0.173 & 0.012 \\
J2R & Intermediate & 12 & -0.171 & -0.331 & -0.160 & 0.202 & 0.257 & 0.016 \\
J2R & J2R &  4 & -0.096 & -0.097 & -0.001 & 0.011 & 0.011 & 0.001 \\
J2R & J2R &  8 & -0.225 & -0.207 & 0.018 & 0.138 & 0.139 & 0.011 \\
J2R & J2R & 12 & -0.171 & -0.153 & 0.018 & 0.170 & 0.170 & 0.014 \\
\bottomrule
\end{tabular}
\begin{flushleft}
\footnotesize
\justifying
Mean complete is the average log rate ratio estimated from the complete data before missingness was imposed, and Mean MI is the average Rubin-pooled estimate after imputation. Mean diff. is the average within-replicate difference between these estimates. Paired SD is the empirical standard deviation of the differences, paired RMSE is their root mean square, and MCSE is the Monte Carlo standard error of the mean difference.
\end{flushleft}
\end{table}
\end{landscape}

\begin{table}[!htbp]
\centering
\caption{Imputation behavior for missing outcomes in the active arm.}
\label{supptab:imputation-behavior}
\scriptsize
\setlength{\tabcolsep}{2pt}
\begin{tabular}{@{}llcccccc@{}}
\toprule
\textbf{Post-trigger outcome mechanism} & \textbf{Imputation rule} & \textbf{Week} & \textbf{True mean} & \textbf{Imputed mean} & \textbf{Bias} & \textbf{True zero prop.} & \textbf{Imputed zero prop.} \\
\midrule
Assigned arm continuation & Assigned arm &  8 & 5.23 & 5.30 & 0.07 & 0.18 & 0.19 \\
Assigned arm continuation & Assigned arm & 12 & 4.01 & 4.22 & 0.21 & 0.22 & 0.24 \\
Assigned arm continuation & Intermediate &  8 & 5.23 & 6.51 & 1.28 & 0.18 & 0.16 \\
Assigned arm continuation & Intermediate & 12 & 4.01 & 5.38 & 1.36 & 0.22 & 0.19 \\
Assigned arm continuation & J2R &  8 & 5.23 & 8.22 & 2.99 & 0.18 & 0.13 \\
Assigned arm continuation & J2R & 12 & 4.01 & 7.13 & 3.12 & 0.22 & 0.16 \\
\addlinespace
Intermediate & Assigned arm &  8 & 6.13 & 5.23 & -0.90 & 0.15 & 0.19 \\
Intermediate & Assigned arm & 12 & 4.98 & 4.16 & -0.82 & 0.18 & 0.23 \\
Intermediate & Intermediate &  8 & 6.13 & 6.29 & 0.16 & 0.15 & 0.16 \\
Intermediate & Intermediate & 12 & 4.98 & 5.11 & 0.13 & 0.18 & 0.19 \\
Intermediate & J2R &  8 & 6.13 & 7.80 & 1.67 & 0.15 & 0.13 \\
Intermediate & J2R & 12 & 4.98 & 6.53 & 1.55 & 0.18 & 0.16 \\
\addlinespace
J2R & Assigned arm &  8 & 7.71 & 5.30 & -2.41 & 0.13 & 0.19 \\
J2R & Assigned arm & 12 & 6.50 & 4.22 & -2.28 & 0.15 & 0.24 \\
J2R & Intermediate &  8 & 7.71 & 6.42 & -1.29 & 0.13 & 0.16 \\
J2R & Intermediate & 12 & 6.50 & 5.23 & -1.27 & 0.15 & 0.19 \\
J2R & J2R &  8 & 7.71 & 7.98 & 0.27 & 0.13 & 0.14 \\
J2R & J2R & 12 & 6.50 & 6.80 & 0.30 & 0.15 & 0.16 \\
\bottomrule
\end{tabular}
\begin{flushleft}
\footnotesize
\justifying
Within each replicate and visit, the true mean and true proportion of zero outcomes were calculated from the complete generated outcomes for participants in the active group whose values were subsequently set to missing. The imputed mean and imputed proportion of zero outcomes were calculated for the same values in each of the 50 completed datasets and then averaged across datasets. The table reports the average of these summaries across included replicates. Bias is the imputed mean minus the true mean. Rows are restricted to missing outcomes for participants in the active group because the reference-based rules differ only for these outcomes from the trigger onward.
\end{flushleft}
\end{table}

\newpage

\section{Application posterior summaries}
\label{supp:application-summaries}

\begin{table}[!htbp]
\centering
\caption{Posterior summaries of selected imputation model parameters and derived conditional daily rates.}
\label{supptab:application-posterior-summary}
\small
\begin{tabularx}{\textwidth}{@{}lXcc@{}}
\toprule
\textbf{Parameter} & \textbf{Interpretation} & \textbf{Estimate} & \textbf{95\% CrI} \\
\midrule
\multicolumn{4}{@{}l}{\textit{Placebo daily rates for the reference covariate profile}}\\
$\exp\left(\eta_{4}\right)$ & Placebo daily rate at Week~4 for the reference covariate profile & 1.720 & (1.444, 2.014) \\
$\exp\left(\eta_{8}\right)$ & Placebo daily rate at Week~8 for the reference covariate profile & 1.543 & (1.293, 1.843) \\
$\exp\left(\eta_{12}\right)$ & Placebo daily rate at Week~12 for the reference covariate profile & 1.453 & (1.181, 1.761) \\
\addlinespace[2pt]
\multicolumn{4}{@{}l}{\textit{Treatment effects}}\\
$\exp\left(\tau_{4}\right)$ & Mirabegron 50~mg versus placebo rate ratio at Week~4 & 0.911 & (0.746, 1.105) \\
$\exp\left(\tau_{8}\right)$ & Mirabegron 50~mg versus placebo rate ratio at Week~8 & 0.854 & (0.694, 1.040) \\
$\exp\left(\tau_{12}\right)$ & Mirabegron 50~mg versus placebo rate ratio at Week~12 & 0.860 & (0.683, 1.088) \\
\addlinespace[2pt]
\multicolumn{4}{@{}l}{\textit{Covariate effects}}\\
$\beta_{\mathrm{base}}$ & Log rate slope for a one-unit increase in centered baseline log rate & 0.762 & (0.669, 0.850) \\
$\zeta_{\mathrm{age}}$ & Log rate ratio: age $\geq 65$ versus $<65$ years & -0.010 & (-0.165, 0.149) \\
$\zeta_{\mathrm{male}}$ & Log rate ratio: male versus female & -0.440 & (-0.671, -0.195) \\
$\zeta_{\mathrm{EE}}$ & Log rate ratio: Eastern Europe versus other regions & -0.164 & (-0.324, -0.010) \\
\addlinespace[2pt]
\multicolumn{4}{@{}l}{\textit{Overdispersion}}\\
$\alpha_{4}$ & NB overdispersion at Week~4 & 1.209 & (1.027, 1.403) \\
$\alpha_{8}$ & NB overdispersion at Week~8 & 1.043 & (0.860, 1.241) \\
$\alpha_{12}$ & NB overdispersion at Week~12 & 0.985 & (0.789, 1.211) \\
\addlinespace[2pt]
\multicolumn{4}{@{}l}{\textit{Dependence}}\\
$\rho$ & AR(1) latent Gaussian correlation between adjacent scheduled visits & 0.464 & (0.396, 0.529) \\
\bottomrule
\end{tabularx}
\begin{flushleft}
\footnotesize
\justifying
Estimate is the posterior mean. CrI = credible interval. Daily rates are per valid diary day and conditional on the reference covariate profile: the sample mean baseline log rate, age $<65$ years, female sex, and a region other than Eastern Europe. Covariate effects are on the log rate scale and common across post-baseline visits.
\end{flushleft}
\end{table}

\newpage

\begin{figure}[!htbp]
\centering
\includegraphics[width=0.68\textwidth]{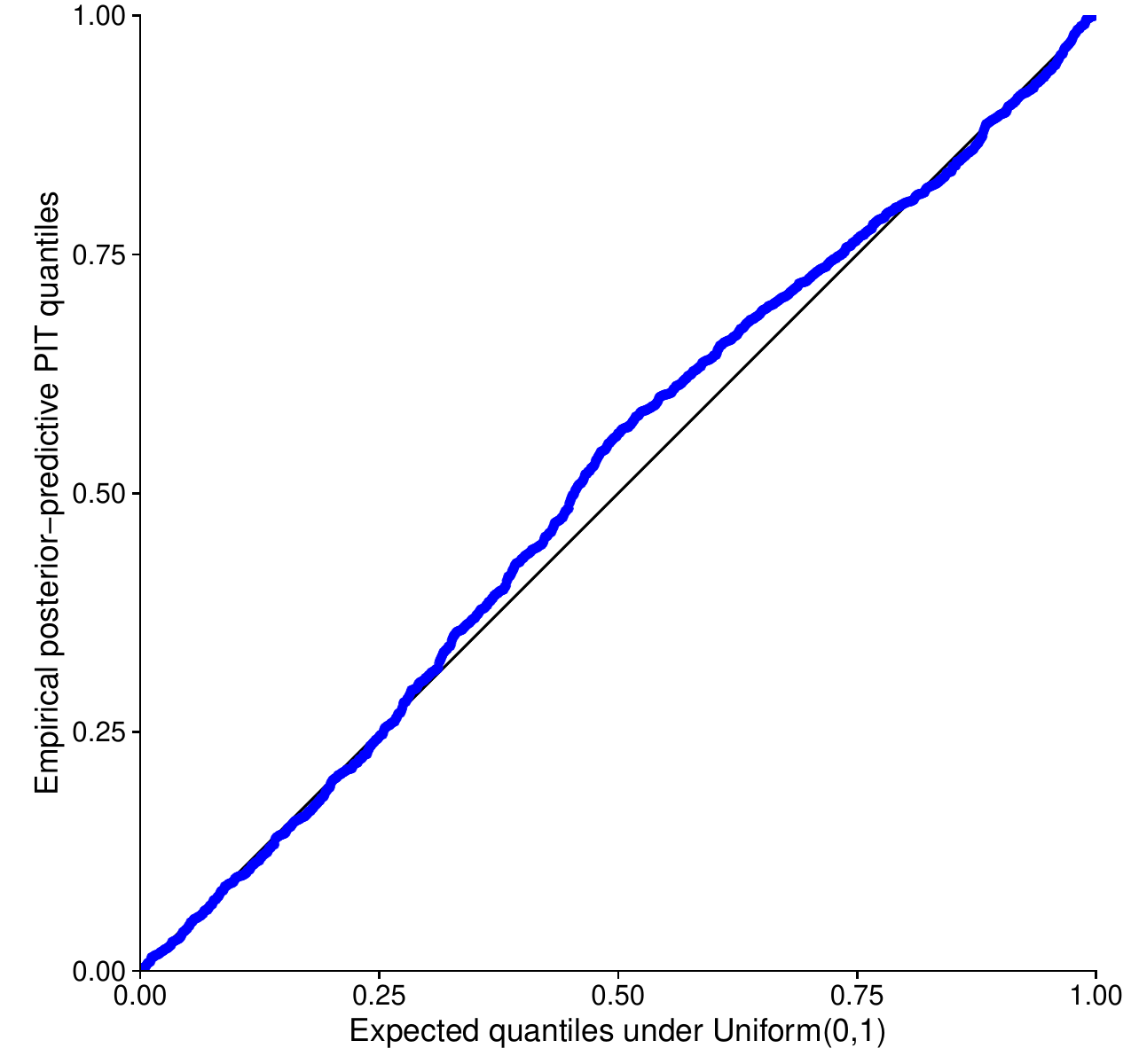}
\caption{
Quantile-quantile (Q-Q) plot of marginal posterior predictive PIT residuals for the fitted NB Gaussian copula model. At each retained posterior draw, one complete post-baseline count trajectory was generated for every participant, conditional on treatment, baseline covariates, and exposure at each visit. Randomized PIT residuals for integer responses were calculated where outcomes were recorded. The points compare empirical residual quantiles with their expected values under a $\mathrm{Uniform}\left(0,1\right)$ distribution; the diagonal represents exact marginal uniformity. Because the simulated trajectories were not conditioned on preceding recorded outcomes, the plot assesses marginal calibration where outcomes were recorded, not the conditional longitudinal dependence.
}
\label{suppfig:application-ppc-residuals}
\end{figure}

\end{document}